\documentclass[smallextended]{svjour3}

\usepackage{cite}
\usepackage[cmex10]{amsmath}
\usepackage{amssymb}
\usepackage{graphics, graphicx, color}
\usepackage{soul}
\usepackage{booktabs}
\usepackage{multirow}
\usepackage{url}
\usepackage{subfigure}
\usepackage{xcolor,soul,framed,caption}
\usepackage{algorithm, algorithmic}
\colorlet{shadecolor}{yellow}

\begin{document}

\title{Study of Network Migration to New Technologies \\ using Agent-based Modeling Techniques}

\author{Tamal Das \and Marek Drogon \and \\ Admela Jukan \and Marco Hoffmann}

\institute{T. Das, M. Drogon, and A. Jukan \at
              Institut f\"{u}r Datentechnik und Kommunikationsnetze \\
              TU Braunschweig \\
              Germany \\
              \email{\{das, drogon, jukan\}@ida.ing.tu-bs.de}           
           \and
           M. Hoffmann \at
              Nokia Siemens Networks \\
              Germany \\
              \email{marco.hoffmann@nsn.com}
}

\maketitle

\begin{abstract} Conventionally, network migration models study competition
between emerging and incumbent technologies by considering the resulting
increase in revenue and associated cost of migration.
We propose to advance the science in the existing network migration models by
considering additional critical factors, including (i) synergistic relationships
across multiple technologies, (ii) reduction in operational expenditures (OpEx)
as a reason to migrate, and, (iii) implications of local network effects on
migration decisions. To this end, we propose a novel agent-based migration model
considering these factors. Based on the model, we analyze the case study of
network migration to two emerging networking paradigms, i.e., IETF Path
Computation Element (PCE) and Software-Defined Networking (SDN). We validate our
model using extensive simulations. Our results demonstrate the synergistic
effects of migration to multiple complementary technologies, and show that a
technology migration may be eased by the joint migration to multiple
technologies. In particular, we find that migration to SDN can be eased by joint
migration to PCE, and that the benefits derived from SDN are best exploited in
combination with PCE, than by itself.

\end{abstract}

\newpage

\section{Introduction}

\par Technical novelties in conjunction with economic factors decide the fate of
an emergent technology, protocol, standard or product in present-day
communication networks. Networks are constantly migrating to new technologies
and services, not only driven by the growth of subscribers base and application
demand, but also new technological advances. The migration is typically a
gradual transition over time, requiring the interoperability and integration of
different network applications, technologies and protocols. For instance, though
the first IPv6 specification was released in 1998 \cite{rfc2460}, the migration
process is still ongoing with only 0.2\% of current Internet traffic being
IPv6-compliant \cite{ddosarbor}. On the other hand, IP backbones today migrate
to router interfaces of a higher capacity at a much faster pace. A typical
carrier IP network is re-planned and increased capacity every 12-18 months, so
that maximum utilization at peak traffic loads is never higher than
approximately 30\%-40\% \cite{headroom}.
Thus, there is no doubt that understanding the strategy and the investments for
network migrations, as well as the expected revenue, network operation expense
and user growth are at the heart of every network migration decision.

\par Technology adoption has been significantly investigated in the literature
using various migration models. However, a few increasingly important factors
have not received enough attention. First, the majority of previous studies
model technology migration in isolation, disregarding the effect of co-existing
technologies in the market. Such studies, thus, do not account for the
synergistic relationships that may exist across technologies, which as a result,
may either facilitate or impede the adoption of a new technology. For instance,
an offering of VPN services with guaranteed QoS may result in a higher revenue,
when combined with automated network management systems. Second, the majority of
migration models are based on the capital expenditures (CapEx) required to
purchase the new technology. However, technology migration often results in
tangible reduction of operational expenditures (OpEx) that is gained over time,
which is typically neglected in the current models. Finally, human decisions are
subject to influence of the social and behavioral factors involved in the
process of migration. For example, although herd mentality (or network effects)
plays a significant role in the adoption of a technology, over and beyond its
technological merits, it is rarely captured in migration models.

\par In this paper, we propose a generic agent-based model to explore network
migration to multiple new \emph{complementary} technologies -- technologies
whose  simultaneous migration is expected to provide greater rewards than the
sum of the rewards derived from their isolated migrations. In addition to CapEx,
our model also incorporates the difference in the OpEx incurred pre- and
post-migration, which significantly affects an agent's decision to migrate. In
the proposed model, an agent also incorporates its estimates of its neighbor's
decision to migrate, in its own migration decision. We accomplish this by means
of both deterministic and probabilistic heuristics. Our results confirm that a
technology migration may be eased by the joint migration of a complementary
technology that is more likely to be adopted.

\par To validate our  proposed model, we analyze the case study of optimal path
computation with joint migration to two emerging networking paradigms, i.e., 
IETF Path Computation Element (PCE \cite{Farrel06}) and Software-Defined
Networking (SDN \cite{SDN-onf-whitepaper}), respectively. The assumed network is
a typical multi-vendor and multi-administration network, where separate
\emph{network islands} of routing systems need to cooperate to provision an
end-to-end connection, and are subject to migration decision pertaining to PCE,
SDN, or both.
PCE enables optimal path computation across network islands, an improved
price/performance ratio, while, at the same time simplifying path computation
operations \cite{metaswitch-whitepaper}. All these benefits added together
attract considerably more users (and in turn traffic) to the network. Exchanges
between PCE and network elements, though standardized, are limited to PCEP
messages, and thus a PCE cannot setup the computed paths itself. To overcome
this limitation, the network operator may decide to migrate to another
technology, say, SDN, which facilitates configuration of all the network
elements, and thereby helps in setting up the computed paths. Moreover,
combining a stateful PCE with OpenFlow provides an efficient solution for
operating transport networks \cite{casellas}.
Thus, there is an implicit correlation between the deployment of PCE and SDN in
a network, which make these two technologies an interesting and practically
relevant case study.

Our paper is organized as follows. Section \ref{sec:rw} discusses the related
literatures and puts our contributions into perspective, while Section
\ref{sec:reference architecture} provides an overview of the technologies that
we study, namely PCE and SDN. Section \ref{sec:model} defines our generic
multi-technology migration model and its application to the case study of
PCE/SDN. Section \ref{sec:rd} discusses the simulation framework to evaluate our
network migration model, and highlights its various aspects using the empirical
results, while, Section \ref{sec:conclusion} presents some concluding remarks.

\section{Related Work and Our Contribution} \label{sec:rw}
In this section, we summarize the previous research in the domain of
our work, and highlight our contributions in this paper.

\subsection{System Dynamics v/s Agent-based Models}
Network migrations are typically studied using \emph{system-dynamics}
\cite{Jin08, Sen10} and \emph{agent-based} models \cite{Macy02, Bonabeau02}.
The former approach is based on aggregate system-wide properties, while, in the
latter approach, simple rules of mutual interaction between agents govern the
evolution of the system. In the system dynamics approach, the migration problem
is treated as a dynamic system in \cite{Jin08, Sen10}, where the rate of
migration depends on the existing number of migrated agents in the system,
according to the traditional diffusion theory of innovation \cite{Bass69}. On
the other hand, in an agent-based approach \cite{Macy02, Bonabeau02}, the system
consists of an ensemble of agents, each trying to increase its own utility.
For example, in \cite{Gill11}, the migration to secure BGP is studied as a
series of decisions by each domain to adopt the technology, based
on the inter-domain routing and the deployment of secure BGP in other domains.
Both approaches demonstrate that the cumulative number of migrations increase
over time assuming a `\emph{S}'-shaped (or sigmoidal) curve, implying that a
majority of migrations is triggered in a short time interval \cite{Borshchev04}.
Despite comparable results, an agent-based approach is preferred over system
dynamics, when the mutual interactions between agents in the system is
non-uniform, for example, when an agent does not interact uniformly with
\emph{all} other agents, but only with those in its local neighborhood. Hence,
we choose agent-based modeling over system dynamics approach for our study in
this paper.

\subsection{Single v/s Multiple Migrations}
\par The network migration problem has typically been studied for a single
technology or protocol (e.g., IPv6 \cite{Joseph07, Trinh10} or secure BGP
\cite{Gill11, Chan06}), where it is assumed that an emerging protocol/technology
\emph{replaces} an incumbent protocol/technology. For example, in case of IPv6,
the models assume that the domain operates either in IPv4 or migrates \emph{fully} to IPv6,
at which point it operates only with IPv6. Even when multiple protocols are
considered, such as S-BGP and soBGP \cite{Chan06}, there is only a single
prevalent protocol, and a decision is made by an agent to adapt to only one of
the competing protocols. Sohn \emph{et al}. propose an economic
evaluation model for a particular aspect of migration, namely, joint development
and standardization of correlated technologies \cite{Sohn11}. Thus, although
majority of the prior migration studies deal with migration of a single technology, the novelty
of our model is in considering multi-technology migrations.

\subsection{CapEx and OpEx considerations}
\par An agent's migration decision is often considered to be solely based on the
CapEx involved. OpEx was recently introduced in cost analysis of migration
research to precisely estimate the cost that the migration to a technology
requires and compare the alternatives \cite{Verbrugge05}. However, the
game-theoretic modeling of migration have not yet considered it \cite{Jin08,
Sen10, Gill11, Chan06, Joseph07, Trinh10}. In this paper, we consider both CapEx
and OpEx in an agent's decision to migrate. In our work, the OpEx reflects an
assumption that the proposed new system will include a level of automation into
the network that alleviates human efforts, resulting in its overall cost
reduction. Our model is thus novel in considering both revenue increase and OpEx
reduction, resulting from migration, as the factors affecting an agent's
decision to migrate.

\begin{figure*}[thb] \centering
\includegraphics[width=1\textwidth]{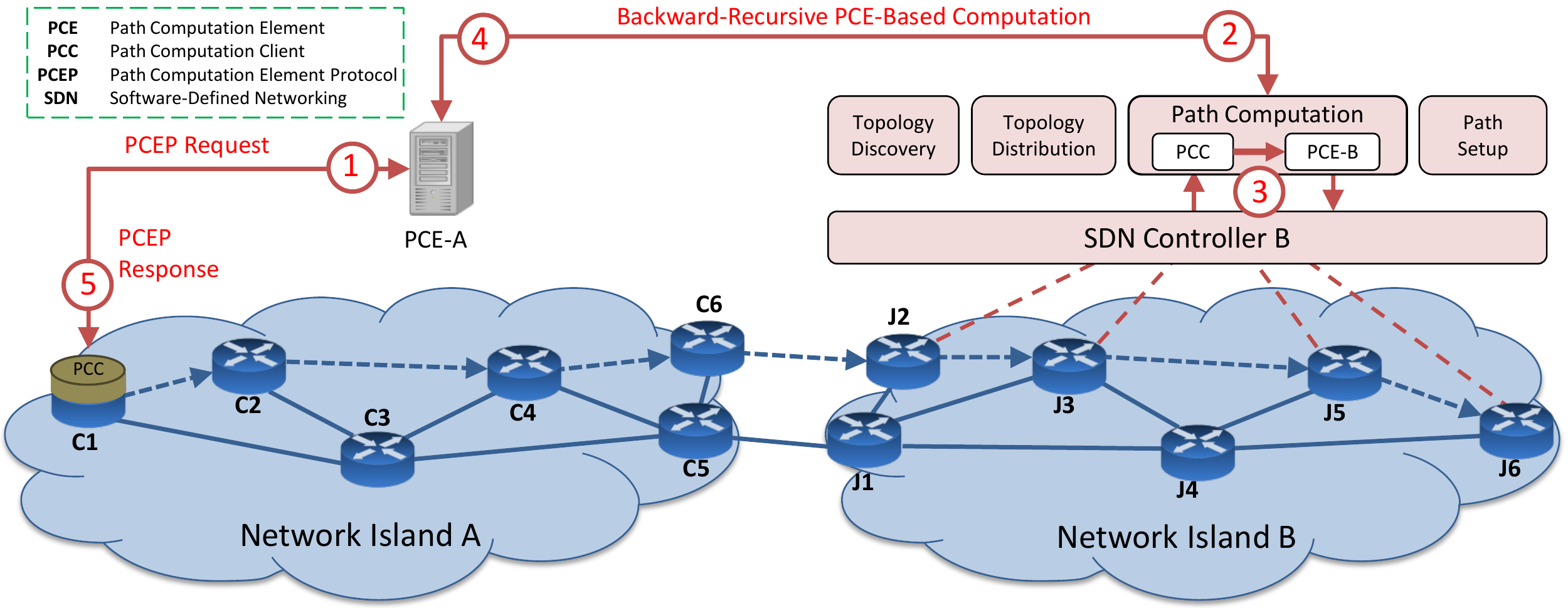}
\caption{Example of connection request setup in a multi-vendor network using
both PCE and SDN.}
\label{figUNI}
\end{figure*}

\subsection{Our Approach and Contribution}
\par This paper extends our previous work on agent-based modeling of network
migration to new technologies \cite{netmig-icc2013}. In this paper, we improve
our CapEx, OpEx and revenue functions used in the network migration model. In
particular, we take into consideration that revenue of a network island follows
economies of scale, i.e.,  every subsequent unit of traffic incurs a lesser cost
to the network operator than the previous. In contrast to \cite{netmig-icc2013},
we differentiate between the OpEx functions in the unmigrated and migrated
states. Unlike \cite{netmig-icc2013}, where only an agent's \emph{immediate}
neighbors were considered to affect the migration choices of the agent in
question, we now extend this effect to include even \emph{distant} neighbors within an
agent's \emph{circle of influence} (defined in Section \ref{sec:estimation
approaches}). The mutual effect of an agent's migration choice on another is
weighted by the reciprocal of the distance between them, restricted to a
threshold distance (beyond which the effect is considered negligible). We also
introduce the notion of \emph{coupling coefficient} to effectively capture the
degree to which two complementary technologies couple with each other. We
propose two novel heuristics for an agent to estimate the strategies of its
neighboring agents in the immediate future, which, in turn, plays a significant
role in the migration decision of the agent in question. Another unique
contribution is in definiton of an agent's
\emph{payoff} from a transition, which is based on its CapEx, OpEx and
revenue functions. An agent migrates only if such a transition results in a
positive payoff for itself.

\par To validate our model, we consider a novel case study of multi-vendor
enterprise network, considering the revenue of a network to vary with the
volume of traffic it transits for its customers. To this end, we consider
simultaneous and correlated deployment of an automated network management system
for path computation (PCE) as well as a programmable network configuration with
SDN controllers, such as based on OpenFlow \cite{openflow}.  We show that the
proposed model is applicable for scenarios, where competing network solutions
(such as multi-vendor environments) collaborate and compete at the same time for
path setup, while aiming at maximum utilization in course of its operation. As
is well-known, inter-operablity of multi-vendor network islands remains
a challenge, and a migration to standardized and programmable automated systems
is an ongoing open problem in carrier networks \cite{ONE}.

\section{Case Study of PCE and SDN: Background and Reference Architecture} \label{sec:reference architecture}

\par In this section, we present an overview of the technologies, namely PCE and
SDN, which we later study using our network migration model. We compare these
two technologies on grounds of path computation and provisioning of a connection
request across multiple network islands in a multi-vendor enterprise network
based on emerging carrier-Ethernet (connection-oriented) networks.

\par \subsection{Technology Overview} PCE is a network-wide centralized server
that receives path computation requests from Path Computation Clients (PCC), and
computes optimal constrained end-to-end paths within a network island.
The PCE can reduce the computation overhead and optimize resource utilization by
computing optimal paths. A major advantage of the PCE architecture is its
ability to compute optimal paths across multiple network islands using the
Backward Recursive Path Computation (BRPC) mechanism \cite{Vasseur09_2}.  In the
BRPC mechanism, PCEs in different islands along a pre-defined chain
progressively compute a Virtual Shortest Path Tree (VSPT) from the destination
to the source, in order to compute the optimal end-to-end path. In absence of
PCE, network islands use Interior Gateway Protocols (like
Open-Shortest-Path-First and Routing-Information-Protocol) and Exterior Gateway
Protocols (like Border Gateway Protocol) to compute paths by means of predefined
routing table entries.

\par SDN is an emerging networking architecture that facilitates programmability
of the network control plane and its separation from the data plane
\cite{SDN-onf-whitepaper}. It provides a centralized control interface to all
the network elements that support SDN protocols, such as Open Flow
\cite{openflow}, which helps in quick experimentation, reconfiguration,
optimization, and monitoring of switching/routing algorithms.
SDN reduces the network OpEx by simplifying operations, optimizing resource
usage through centralized data/algorithms, and simplifying network software
upgrades. SDN also significantly cuts down a network operator's CapEx, since a
commercial-off-the-shelf (COTS) server with a high-end CPU is much cheaper than a high-end router
\cite{metaswitch-whitepaper}. Further, SDN offers the possibilities of dynamic
network topologies and network virtualization, which makes it currently a highly
popular paradigm \cite{google-openflow}.

\par \subsection{Reference Architecture}

\par Figure \ref{figUNI} illustrates an automated connection setup in a typical
multi-vendor, multi-technology network island setting. Two different network
islands are shown. The network island \emph{A} consists of six different IP
routers (C1-C6) from vendor C (e.g. Cisco), whereas, the network island \emph{B} consists
of six IP routers (J1-J6) from Vendor J (e.g. Juniper).

\par The choice of technology for network island \emph{A} is PCE-only. A Path
Computation Element (PCE-A) is used within the network to compute constrained-based
paths across intra- and inter-network island scenarios. The topology discovery
and distribution is handled via separate protocols, such as OSPF, and the
RSVP-TE protocol can be used for path setup. All protocols need to be installed and
configured separately on every router, with only limited
possibilities for functionality extensions and optimizations. Network island
\emph{A} has the possibility to migrate to SDN in future.
The migration to SDN would benefit network island \emph{A} by introducing a central intelligence that is
capable of automating processes, thus saving OpEx.

\par The choice of technology for network island \emph{B} is PCE+SDN. In this
network, a central intelligence (SDN Controller B) is directly accessing every
router in the network, via a SDN router interface for flexible
configuration of router equipment. The SDN Controller \emph{B} can choose from
different network functionalities, such as topology discovery, topology
distribution, path computation and path setup. All functionalities are
software-defined modules, that are programmed on top of the SDN Controller for
on-the-fly functionality extensions and optimizations. Network Island \emph{B} already
has the maximal technology set of our case-study. All operations can be fully
automated, thus no manual intervention is necessary, resulting in significant
OpEx savings.

\par Both network island are connected via two inter-network island connections.
A path computation request from C1 to J6 is handled via the PCEP protocol
supported by both network islands. Router C1 sends a PCEP Request message to
PCE-A (1). PCE-A tries to compute an end-to-end path to J6, but does not have
enough information to calculate this path. PCE-A knows the existence of PCE-B
(either through pre-configuration or discovery), and issues a Backward-Recursive
PCE-Based Computation. The PCE-B computes the shortest path from J2 to J6 by
accessing the SDN Controller B, that is retrieving all necessary information
from the Topology Discovery and Distribution for optimal path computation within
network island \emph{B}. The optimal path from the entry-router (J2) to the destination
(J6) is returned to PCE-A (4). PCE-A now has the optimal path from J2 to J6 and
computes the best path from C1 to J2 and returns the whole path to C1 (5). The
resulting path (C1-C2-C4-C6-J2-J3-J5-J6) is used to reach the
destination.

\par A couple of comments are worth noting.
First, although each PCE sees only its own network topology, BRPC enables an
optimized (i.e., best QoS) end-to-end path. Second, despite the fact that each
SDN controller can implement its own path computation algorithm, the assumption
here is that they often tend to be highly proprietary in nature. Thus, lack of
standards makes it hard for SDNs to interoperate in a multi-vendor setting ---
that is where the IETF-standardized approach with PCE comes in as an effective
solution for interoperability.

\subsection{Interplay involved in joint migration to PCE and SDN} As can be
seen, the interplay involved in joint migration to PCE and SDN can lead to
interesting, non-trivial network behavior, which we now discuss in further
detail.

\par In our analysis, we assume a typical control plane with management network
control environment. A network operator has an advantage in migrating to SDN
over PCE, as a PCE can only compute paths, while a SDN controller can as well
provision the computed paths in a highly programmable fashion. However, as
previously mentioned, in a typical multi-vendor setting, a PCE has advantages
over SDN. This is because PCE (being standardized) can communicate with
neighboring PCEs, whereas, SDNs (being non-standardized) cannot. Thus, larger
the diversity of network equipment in the same network, greater is the incentive
for the network operator to migrate to PCE than SDN, on account of
interoperability considerations.

\par Within a network island, a SDN controller is \emph{likely} to be able to
provision a path, even when a PCE may not. A typical SDN controller, based on
OpenFlow, is in fact expected to access and configure network elements at the
operator's liking, including the handling of lower layers of the network, such
as optical circuits. Not only can a SDN controller find paths that a PCE is
requesting, but it can potentially even reconfigure the whole network such that
a totally \emph{new} path is configured to provision a connection request. Thus,
SDN can potentially create paths with a better QoS unlike PCE, which only
computes paths based on requests. Hence, the end-user benefits more if its
network provider migrates to SDN, than PCE. On the other hand, as the PCE
protocol is reactive in nature, unlike SDN (which is proactive), end-users stand
to gain more from PCE than from SDN.

\par Whereas a SDN controller is triggered by the NMS/OSS in the network, PCE
can be triggered by the end-user. Both SDN and PCE benefit the network operator
through OpEx reduction; whereas, PCE, in addition, benefits the end-user by
providing improved QoS for end-to-end connections involving multiple vendors.
Although a network does not attract any additional traffic by migrating to
PCE/SDN, it benefits significantly by reducing its OpEx after migration.

\par As SDN offers more functionalities than PCE (such as path provisioning,
topology discovery and topology distribution), both the CapEx required to
migrate to SDN and the resulting OpEx is more than that required to migrate to
PCE. In addition, unlike PCE, the non-standardized nature of SDN adds to its
OpEx.
Further, the CapEx involved in simultaneous migration of a network island to PCE
and SDN is less than the sum of the CapEx involved in separate migrations to PCE
and SDN. This is because, in case of simultaneous migrations, the PCE can be
incorporated \emph{within} the SDN controller, thus providing an integrated
platform at a reduced cost.

\par In summary, network islands that migrate to PCE can compute optimal paths
(i.e., with QoS), which can be provisioned using automated network management
frameworks, such as SDN. Thus, it is clear that SDN controllers, with its reach
limited to a network island, ideally complement the PCEs that can communicate
across networks, thereby, enabling optimal end-to-end, multi-vendor,
multi-domain path computation and provisioning under QoS constraints.

\section {Multi-Technology Network Migration Model} \label{sec:model}

\par In this Section, we present our generic agent-based model for studying
network migration to complementary technologies. As a case study, we apply our
model to study the dynamics of joint migration to multi-vendor path computation
and provisioning, namely PCE and SDN, respectively.

\subsection{Generic Model}

\par Our model captures the collaborative and competitive business relationships
between the agents and also the inter-dependencies involved in their
decision-making process. The time is discretized, and thus the model progresses
in time-steps. The agents are considered to be \emph{myopic} (in time) in their
decision-making and are assumed to act under \emph{complete information}. The
former assumption entails each agent optimizing their strategy choices
\emph{locally} (in time), while the latter means that each agent is
aware of the complete network topology as well as the past strategy choices of
all other agents.

\par \emph{Notations}: The agents in our model are denoted by $N_1, N_2, \cdots, N_i,
\cdots$. An agent's strategy set is represented by a compatible combination of
the available strategies. We denote this universal set of strategies available
for the agents to choose from, by two sets of \emph{substitutive} strategies, $S
= \left\{ S_{u}, S_{v}\right\}$, where $u$ and $v$ are the \emph{complementary}
technologies under consideration, which implies that the payoff that an agent
derives by adopting both of them simultaneously is higher than the sum of its
payoffs derived by adopting each of them separately (while, no such
relationship is assumed to exist between $s_{u,0}$ and $s_{v,0}$). Here,
$S_{u}=\left\{s_{u,0},s_{u,1}\right\}$ represents the strategy of non-adoption
and adoption of technology $u$, respectively. Similarly,
$S_{v}=\left\{s_{v,0},s_{v,1}\right\}$ represents the strategy of non-adoption
and adoption of technology $v$, respectively. Further, $s_{u,0}$ (or $s_{v,0}$) and
$s_{u,1}$ (or $s_{v,1}$) are \emph{substitutive} strategies, as an agent can
adopt only one of them at any given time. Thus, an agent's \emph{strategy
set} for any given time-step is denoted by $a=\left\{s_{u,
k_1},s_{v,k_2}\right\}$, where, $k_1,k_2 \in \{0,1\}$. The volume of sales of
agent $N_i$ given its strategy set $a$ is denoted by $T_{a}^i$.

\par An agent's revenue and OpEx depends on its amount of sales, while the cost of
changing its strategy set depends on the required CapEx. Considering
this, we define the following notations.
\begin{displaymath}
\begin{array}{rcl}
C_i(a \to a^\prime) & \triangleq & \textrm{CapEx of } N_i \textrm{ to migrate from } a \textrm{ to } a^\prime  \\
R_i(a) & \triangleq & \textrm{Revenue of } N_i \textrm{ with strategy set } a\\
O_i(a) & \triangleq & \textrm{OpEx of } N_i \textrm{ with strategy set } a
\end{array}
\end{displaymath}
where, $a$ denotes the current strategy set of agent $N_i$ and $a^\prime$
denotes the strategy set to which $N_i$ migrates in the subsequent
time-step. We define the payoff of an agent on migrating to a different strategy set by the \emph{return on investment} it derives from such a decision. The payoff derived by an agent on migrating from $a$ to
$a^\prime$ is thus given by the CapEx involved in the migration and the
corresponding change in revenue and OpEx as:

\begin{multline}\label{eq:genericPayoffDef}
P_i({a \to a^\prime}) = \frac{\Delta\textrm{(Revenue)} - \left[\textrm{CapEx} + \Delta\textrm{(OpEx)}\right] }{\textrm{CapEx}}\\
=\frac{ \left[R_i(a^\prime)-R_i(a)\right] - C_i(a \to a^\prime) -\left[O_i(a^\prime)-O_i(a)\right]}{C_i(a \to a^\prime)}
\end{multline}

\par Each agent thus optimizes its strategy choices at every time-step based on
its payoff maximization in the immediate future. Note that each of the CapEx,
OpEx and revenue functions, in turn depend on the amount of sales of agent
$N_i$, namely, $T_{a}^i$ and $T_{a^\prime}^i$. $T_{a}^i$, viz. the current
amount of sales of agent $N_i$, can be deterministically computed by $N_i$ from
its system measurements, whereas, $T_{a^\prime}^i$, viz. the expected
amount of sales of $N_i$ on transitioning from strategy set $a$ to $a^\prime$,
is unknown. We next present two different approaches to estimate this expected
amount of sales, $T_{a^\prime}^i$.

\subsection{Estimation of $T_{a^\prime}^i$} \label{sec:estimation approaches}
The amount of sales of an agent primarily depends on the agent's technology
choices, which in turn is significantly affected by the strategy choices of the
neighboring agents within its `\emph{circle of influence}'. We define this
novel concept referred to as a \emph{circle of influence} of an agent as its
neighborhood comprising of all agents, whose technology choices
\emph{significantly} affects the migration decision of the agent under
consideration. In other words, we capture the notion of \emph{local network
effects} \cite{localNetworkEffects} using our concept of circle of influence.
Thus, the circle of influence of, say, agent $N_i$ comprises of all agents whose
distance from agent $N_i$ is bounded by a threshold distance (by the shortest
path), say $\delta_i$.
We call $\delta_i$ as the `\emph{relevant radius}' of $N_i$'s circle of
influence. We also note that the mutual effect of the strategy choices of two
agents (within each others circle of influence) is inversely proportional to the
distance between them. To capture this aspect, we define the \emph{effective
migration coefficient} of agent $N_i$, as the weighted average of the strategy
sets of all agents within $N_i$'s circle of influence; the weights being the
reciprocal of the distance of the corresponding agent from $N_i$.
The influence of the strategy choices of an agent, which does not fall within
$N_i$'s circle of influence, on $N_i$'s migration decision is, hence, considered
negligible. Thus, for an agent to estimate its expected amount of sales in the
immediate future, it needs to estimate of the strategy choices of all agents
within its circle of influence, in the immediate future. This computation of
effective migration coefficient for agent $N_i$ is further illustrated in Algorithm \ref{pseudocode}.

\begin{algorithm}
\caption{Effective migration coefficient of agent $N_i$}
\label{pseudocode}
\begin{algorithmic}
\STATE {$num\gets$ 0}
\STATE {$den\gets$ 0}
\FORALL{agents $N_j$}
	\IF {$i \neq j$}
	\STATE {$dist\gets$ minumum number of hops between $N_i$ and $N_j$}
		\IF {$dist<\delta_i$}
			\STATE {$num \gets num+\dfrac{\textrm{Migration state of $N_j$}}{dist}$}
			\STATE {$den \gets den+\dfrac{1}{dist}$}
		\ENDIF
	\ENDIF
\ENDFOR
\STATE {effective migration coefficient of $N_i\gets \dfrac{num}{den}$}
\end{algorithmic}
\end{algorithm}

\begin{figure}
\begin{center}
\includegraphics[width=0.9\textwidth]{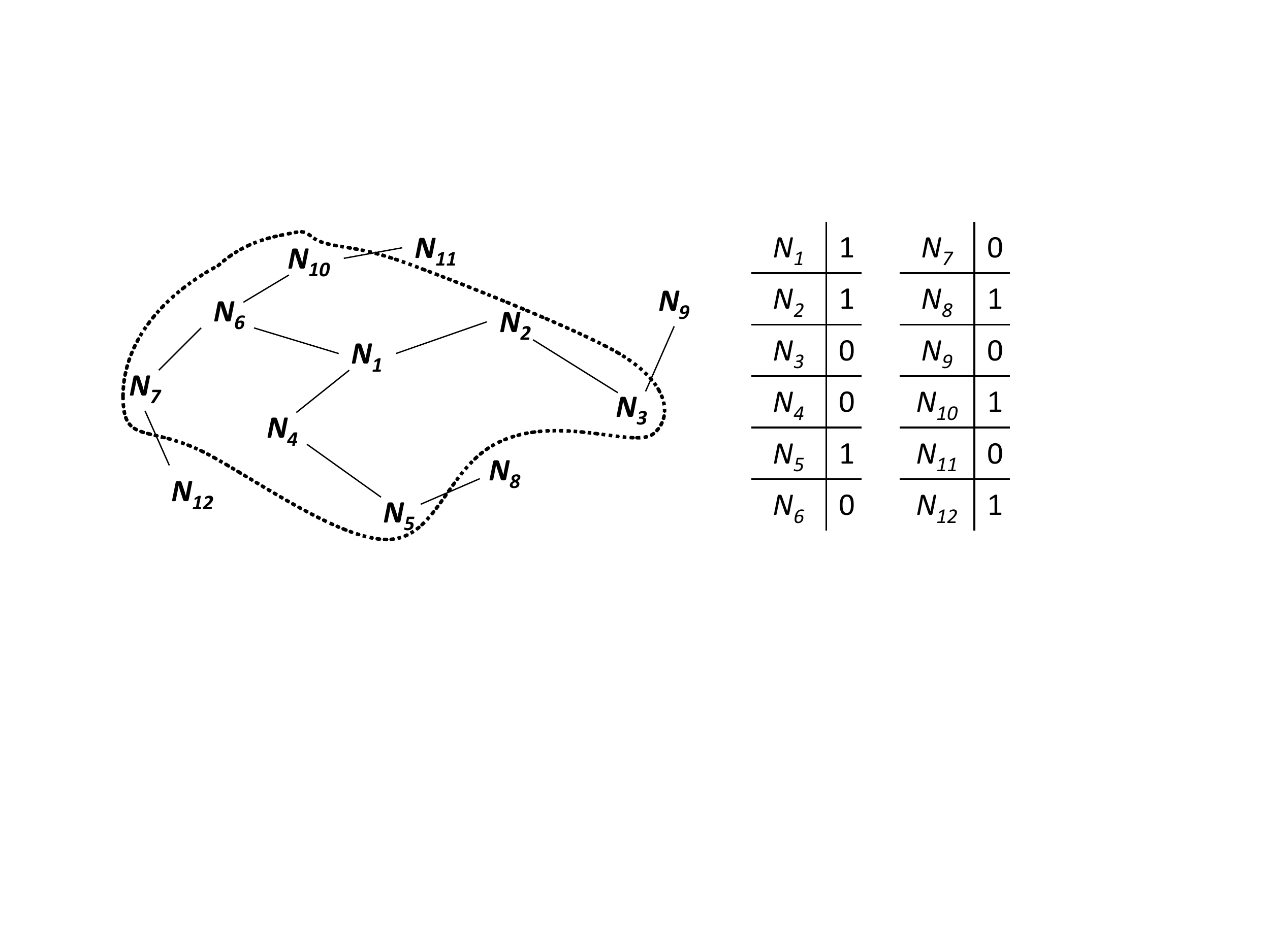}
\caption{Circle of influence}
\label{fig:effMigCoeff example}
\end{center}
\end{figure}

Figure \ref{fig:effMigCoeff example} shows a 12-node network to illustrate the
above mentioned concepts. In this topology, the relevant radius of agent $N_1$,
i.e. $\delta_1$, is considered to be 2 hops, and $N_1$'s circle of influence
is marked by a dotted line. The adjoinging tables in Figure
\ref{fig:effMigCoeff example} list the current migration state of all agents
in the network. Given this, the effective migration coefficient of $N_1$ is
thus given by,

$$
\dfrac{\overbrace{\dfrac{1}{1}}^{N_2}+\overbrace{\dfrac{0}{2}}^{N_3}+\overbrace{\dfrac{0}{1}}^{N_4}+\overbrace{\dfrac{1}{2}}^{N_5}+\overbrace{\dfrac{0}{1}}^{N_6}+\overbrace{\dfrac{0}{2}}^{N_7}+\overbrace{\dfrac{1}{2}}^{N_{10}}}{\dfrac{1}{1}+\dfrac{1}{2}+\dfrac{1}{1}+\dfrac{1}{2}+\dfrac{1}{1}+\dfrac{1}{2}+\dfrac{1}{2}}=\dfrac{2}{5}=0.4
$$

\par We next present two
heuristics for an agent to estimate its neighbor's strategy in the subsequent
time-slot, based on \emph{probabilistic} and \emph{deterministic} methods. The
underlying rationale behind both these heuristics is that an agent's strategy
choice is very likely to vary with that of the majority of the agents in its
circle of influence.

\subsubsection{Deterministic Strategy Estimation} In the \emph{deterministic approach}, an
agent considers the strategy choices of its neighboring agents to be the same as
that of the majority of the agents in their circle of influence. Thus, while
agent $N_i$ is estimating its future amount of sales, if $N_j$ is within $N_i$'s
circle of influence, and if more than 50\% of the agents in $N_j$'s circle of
influence employ strategy set $a$ in the current time-step, then $N_i$ expects
$N_j$ to switch to strategy set $a$ in the next time-step, under this
approach.

\subsubsection{Probabilistic Strategy Estimation} In this estimation approach,
an agent considers the \emph{probability} of its neighbor's strategy choice in
the subsequent time-slot to be $a$, as $x$, if $x$ denotes the fraction of
agents with strategy set $a$, in this neighbor's circle of influence, in the
current time-slot. Thus, in the process of agent $N_i$ estimating its future
amount of sales, if $N_j$ is within $N_i$'s circle of influence, and if, say,
30\% of the agents in $N_j$'s circle of influence employ strategy set $a$ in the
current time-step, then $N_i$ assumes the probability of $N_j$ switching its
strategy set to $a$ in the subsequent time-step as 0.3, under this approach.

\par Note that it is due to our assumption of complete information that
these heuristics can be realized. Figure \ref{fig:deterministic probabilistic
curves} plots the probability of migration of an agent using deterministic
and probabilistic estimation approaches, as a function of its effective
migration coefficient.

\par An agent thus estimates the strategy sets of all agents within its circle
of influence in the immediate future, using one of the two strategy estimation
approaches, mentioned above. It thus disregards the future strategy choices of
agents outside its circle of influence, and assumes them to maintain the same
strategy set, in the subsequent time-step. Thereafter, the agent takes note of
its own set of possible transitions from its current state, i.e., $a \to \{a_1,
a_2, \ldots\}$ (see Figure \ref{fig:StrategySetTransitions}). It then computes
the payoffs resulting from each of its possible transitions, in sync with the
strategy set estimations of the agents within its circle of influence, i.e., 
$\{P_i(a \to a_j)\},\forall j$, and accordingly chooses its future strategy set
as the one that maximizes its resulting payoff, i.e., $a^\prime =
\operatorname{arg\,max}_{a_j} \{P_i(a \to a_j)\}$, given its current strategy
set $a$.
In this way, an agent optimizes its strategy set at each time-step.

\begin{figure}
\begin{center}
\includegraphics[width=6cm]{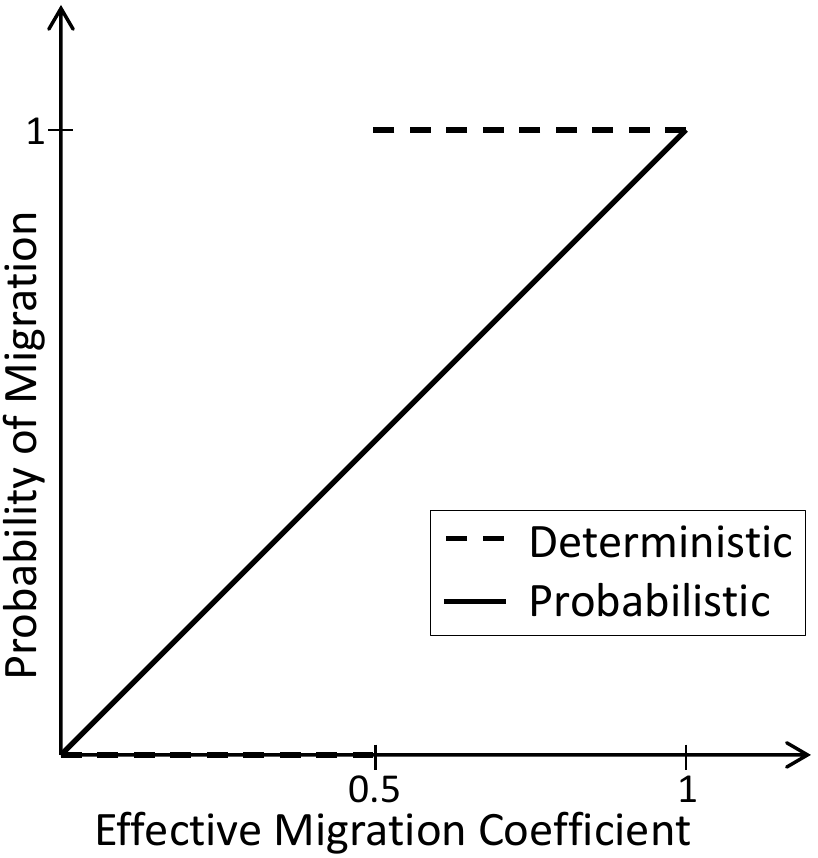}
\caption{Probability of Migration of an agent in deterministic and
probabilistic estimation approaches, as a function of its effective migration
coefficient}
\label{fig:deterministic probabilistic curves}
\end{center}
\end{figure}

\subsection{Agent-based Model Applied} In this subsection, we customize our
generic network migration to the particular scenario of migration to
PCE and SDN.

\begin{table}[ht]
\centering 
\begin{tabular}{c c c} 
��������Agent��������� &	$\longleftrightarrow$	& 	Network Island��� \\
��������Strategy��� &	$\longleftrightarrow$	& 	 Technology Choice \\
��������Amount of Sales &	$\longleftrightarrow$	& 	 Amount of Traffic� \\
��������Technology $u$��� &	$\longleftrightarrow$	& 	 PCE��������������� \\
��������Technology $v$��� &	$\longleftrightarrow$	& 	 SDN���������������
\end{tabular}
\caption{Mapping generic migration model to PCE/SDN} 
\label{table: mapping generic model to pce/sdn} 
\end{table}

\par Table \ref{table: mapping generic model to pce/sdn} summarizes the mappings
between the generic network migration model and PCE/SDN scenario. In the context
of PCE/SDN, agents translate to network islands, strategies correspond to
technology choices, amount of sales relate to the amount of traffic that a
network transits for its customers, technology $u$ maps to PCE, while,
technology $v$ maps to SDN.

\par
\begin{figure} [!hbt]
\begin{center}
\includegraphics[width=0.6\textwidth]{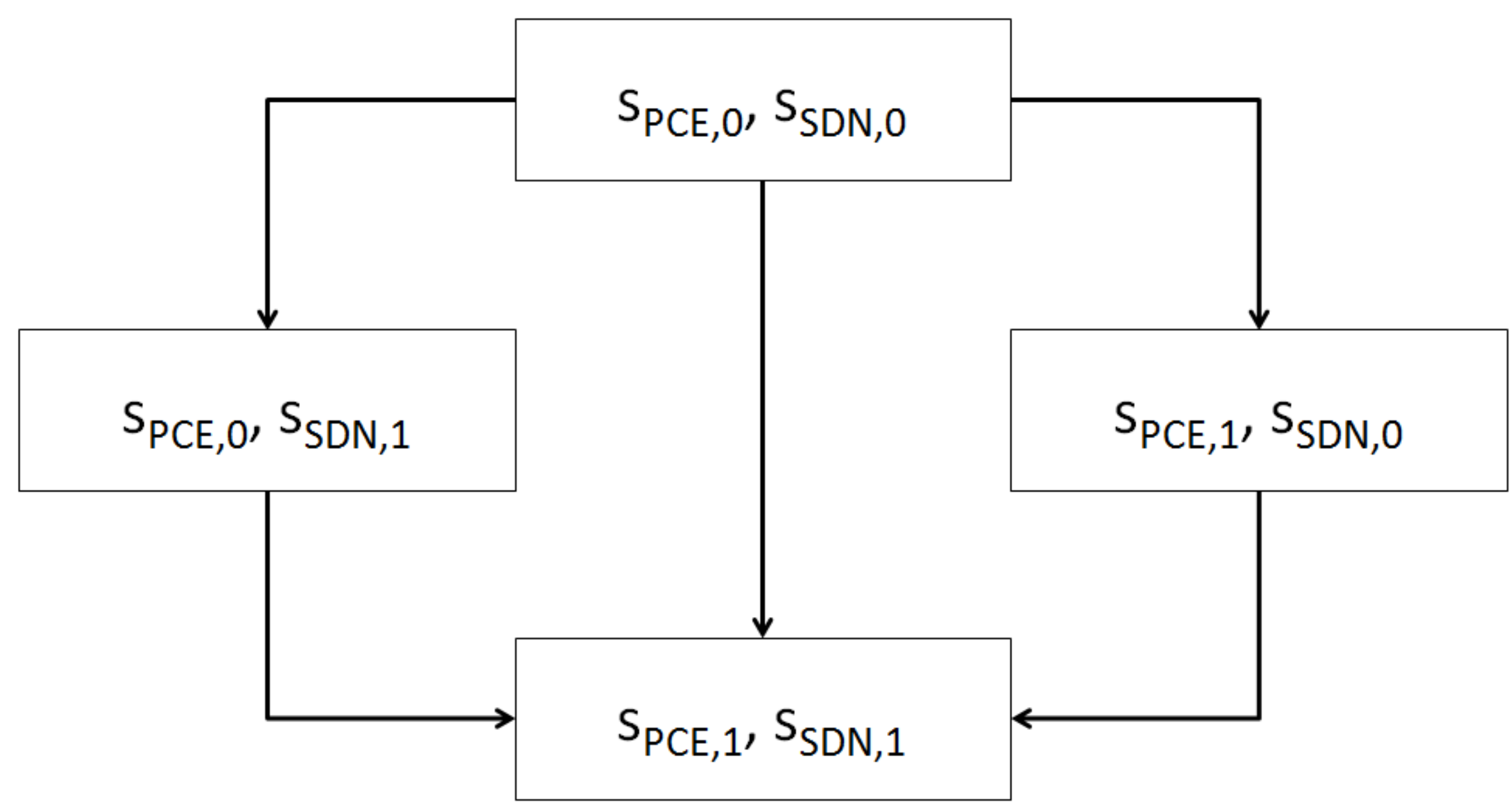}
\label{fig:str}
\end{center}
\caption{Strategy set transitions in a network.}
\label{fig:StrategySetTransitions}
\end{figure}

 Figure \ref{fig:StrategySetTransitions} shows all
possible strategy set transitions for a network island, under the assumption that an
island that has once migrated to $s_{\textrm{PCE},1}$ or $s_{\textrm{SDN},1}$
does not revert back to $s_{\textrm{PCE},0}$ or $s_{\textrm{SDN},0}$,
respectively, in the future. This assumption is justified because the
functionalities provided by PCE and SDN are beneficial to a network, irrespective
of external factors, such as the technology choices of other network islands,
etc.
For instance, a migrated node definitely saves its OpEx, even if the resulting
traffic does not increase post-migration (see Figure \ref{fig:cause of migration}).

\par A network island incurs CapEx if it migrates to PCE or SDN.
Secondly, the CapEx of a network island is expected to follow economies of
scale, i.e.,  every subsequent unit of traffic incurs a lesser CapEx than the
previous.
We, hence, assume the CapEx to vary with the square root of the network traffic.
In addition, on account of the complementary relationship between PCE and SDN,
the CapEx incurred by a network island in migrating to both the technologies
simultaneously is less than the sum of the CapEx incurred by migrating to each
of them separately. This is because, although PCE and SDN are separate
components, if a network island migrates to both of them simultaneously, it can
integrate both the technologies into a single, integrated component, leading to
a reduced CapEx, as compared to a PCE component, and a separate SDN component.
Considering both these aspects, the CapEx of network island $N_i$ from the
generic model in equation \eqref{eq:genericPayoffDef} can be expressed, in this case, as
\begin{equation}\label{eq:capex definition}
C_i(a \to a^\prime) = c_i(a,a^\prime)\sqrt{T_{a^\prime}^i}
\end{equation}
where, $c_i(a,a^\prime) \in [0,1]$ is a coefficient given by,
\begin{multline} \label{eq:capex coefficients}
c_i(a,a^\prime)=
\left\{\begin{array}{c l}
c_\textrm{PCE}& \{s_{\textrm{PCE},0},s_{\textrm{SDN},k}\}\to\{s_{\textrm{PCE},1},s_{\textrm{SDN},k}\} \\
c_\textrm{SDN}& \{s_{\textrm{PCE},k},s_{\textrm{SDN},0}\}\to\{s_{\textrm{PCE},k},s_{\textrm{SDN},1}\} \\
\left(\dfrac{c_\textrm{PCE}+c_\textrm{SDN}}{\eta}\right)&\{s_{\textrm{PCE},0},s_{\textrm{SDN},0}\} \to \{s_{\textrm{PCE},1},s_{\textrm{SDN},1}\} \\
\end{array} \right.
\end{multline}
where, $k \in \{0,1\}, c_\textrm{PCE}, c_\textrm{SDN} \in [0,1]$ and $\eta \in [1,2]$ denotes the \emph{coupling coefficient} ---
$\eta = 1$ implies fully independent technologies, such that, migrating to
both these technologies simultaneously is equivalent to migrating to each of
them separately, whereas, $\eta = 2$ implies fully substitutive
technologies, such that, migrating to both of them simultaneously is equivalent to
migrating to any one of them. In the context of PCE and SDN, we consider
$\eta = 1.5$ in this paper.

\par The revenue of a network island primarily depends on the amount of traffic
flowing through it, and does not vary with the set of technologies deployed by
the network operator. This is because the revenue comes from the customer, who
is oblivious to the technology adopted by its network operator. The customer,
generally, pays to the network operator, solely based on the amount of traffic
that the operator transits for it. In addition, revenue of a network island is
expected to follow economies of scale.
We, thus, consider the revenue of a network island to vary as the square of the
network traffic.
And, given the \emph{qualitative} nature of our model, without loss of
generality, we set,
\begin{equation} \label{eq:revenue definition}
R_i(a)=(T_{a}^i)^2
\end{equation}

Similar to CapEx, the OpEx of PCE and SDN in a network island is expected to
follow economies of scale, i.e.,  every subsequent unit of traffic incurs a lesser
CapEx than the previous. Hence, we consider the OpEx of a network island to vary
with the square root of the network traffic. Thus,
\begin{equation} \label{eq:opex definition}
O_i(a) = \alpha_i(a)\sqrt{T_a^i}\\
\end{equation}
where, $\alpha_i(a)$ is a coefficient given by,
\begin{equation} \label{eq:opex coefficients}
\alpha_i(a)=
\left\{\begin{array}{c l}
\alpha_{\overline{\textrm{PCE}}}+\alpha_{\overline{\textrm{SDN}}} & a = \{s_\textrm{PCE,0},s_\textrm{SDN,0}\} \\
\alpha_{\overline{\textrm{PCE}}}+\alpha_\textrm{SDN} & a = \{s_\textrm{PCE,0},s_\textrm{SDN,1}\} \\
\alpha_\textrm{PCE}+\alpha_{\overline{\textrm{SDN}}} & a = \{s_\textrm{PCE,1},s_\textrm{SDN,0}\} \\
\left(\dfrac{\alpha_\textrm{PCE}+\alpha_\textrm{SDN}}{\eta}\right) & a = \{s_\textrm{PCE,1},s_\textrm{SDN,1}\} \\
\end{array} \right.
\end{equation}
where, the overline operator $(\overline{\textrm{PCE}} \textrm{ and }
\overline{\textrm{SDN}})$ denotes the alternatives available (say, manual
operations) to the corresponding
technology, (i.e.,  PCE and SDN, respectively) and $\alpha_{\overline{\textrm{PCE}}}, \alpha_{\overline{\textrm{SDN}}}, \alpha_\textrm{PCE}, \alpha_\textrm{SDN} \in [0,1]$. Thus, $\alpha_\textrm{PCE}$ is the coefficient of
the PCE component of OpEx in the presence of PCE, whereas,
$\alpha_{\overline{\textrm{PCE}}}$ denotes the corresponding coefficient in the
absence of PCE. Similarly, for $\alpha_{\textrm{SDN}}$ and
$\alpha_{\overline{\textrm{SDN}}}$. The presence of $\eta$ in equation
\eqref{eq:opex coefficients} captures the complementary relationship between
PCE and SDN, i.e.,  the OpEx incurred by a network
island on migrating to both the technologies simultaneously is
less than the sum of the OpEx incurred by migrating to each
of them separately.

\par We also note that both PCE and SDN are significantly more efficient than
their alternative technologies (say, manual operations). Thus, a domain
migrating to either PCE or SDN is expected to result in a non-negative change in
OpEx, or in other words, in OpEx savings. To put it mathematically, the
corresponding OpEx coefficients of PCE and SDN, pre- and post-migration must
satisfy the following inequalities.
\begin{equation} \label{eq:unmig > mig}
\begin{array}{c}
\alpha_\textrm{PCE} < \alpha_{\overline{\textrm{PCE}}} \\
\alpha_\textrm{SDN} < \alpha_{\overline{\textrm{SDN}}}
\end{array}
\end{equation}

\par In all migration scenarios in general, and in migration to  to PCE or SDN
in particular, the major investment is often in the CapEx involved, whereas, the post-migration
OpEx decreases, compared to pre-migration OpEx costs. Moreover, the CapEx of
migration generally supersedes the post-migration OpEx costs by a significant margin.
This, in conjunction with equations \eqref{eq:capex coefficients} and
\eqref{eq:opex coefficients}, leads us to state,

\begin{equation} \label{eq:capex>opex, pce}
c_\textrm{PCE} > \textrm{max}\left\{\alpha_\textrm{PCE}+\alpha_{\overline{\textrm{SDN}}} , \dfrac{\alpha_\textrm{PCE}+\alpha_\textrm{SDN}}{\eta}\right\}
\end{equation}

\begin{equation} \label{eq:capex>opex, sdn}
c_\textrm{SDN} > \textrm{max}\left\{\alpha_{\overline{\textrm{PCE}}}+\alpha_\textrm{SDN} , \dfrac{\alpha_\textrm{PCE}+\alpha_\textrm{SDN}}{\eta}\right\}
\end{equation}

\begin{equation} \label{eq:capex>opex, both}
\dfrac{c_\textrm{PCE}+c_\textrm{SDN}}{\eta} > \dfrac{\alpha_\textrm{PCE}+\alpha_\textrm{SDN}}{\eta}
\end{equation}

Equation \eqref{eq:capex>opex, pce} results from the fact that the CapEx
of migrating from
$\{s_{\textrm{PCE},0},s_{\textrm{SDN},k}\}$
to $\{s_{\textrm{PCE},1},s_{\textrm{SDN},k}\}$ is greater than the
post-migratin OpEx costs in both cases ($k=0,1$). However, since the
CapEx and OpEx functions are similar in nature, this relationship also
holds for the corresponding coefficients. Thus, the corresponding CapEx
coefficient ($c_\textrm{PCE}$) must be greater than both the OpEx coefficients
in the two scenarios (viz.,
$\alpha_\textrm{PCE}+\alpha_{\overline{\textrm{SDN}}}$ and
$\dfrac{\alpha_\textrm{PCE}+\alpha_\textrm{SDN}}{\eta}$). Equations
\eqref{eq:capex>opex, sdn} and \eqref{eq:capex>opex, both} result from similar
arguments for migrations from
$\{s_{\textrm{PCE},k},s_{\textrm{SDN},0}\}$
to $\{s_{\textrm{PCE},k},s_{\textrm{SDN},1}\}$, and from
$\{s_{\textrm{PCE},0},s_{\textrm{SDN},0}\}$
to $\{s_{\textrm{PCE},1},s_{\textrm{SDN},1}\}$, respectively.

\par Eliminating $\alpha_{\overline{\textrm{SDN}}}$ between equations
\eqref{eq:unmig > mig} and \eqref{eq:capex>opex, pce}, we have,
\begin{equation}
c_\textrm{PCE} > \alpha_\textrm{PCE}+\alpha_\textrm{SDN}
\end{equation}

Similarly, eliminating $\alpha_{\overline{\textrm{PCE}}}$ between equations
\eqref{eq:unmig > mig} and \eqref{eq:capex>opex, sdn}, we have,
\begin{equation} \label{eq:constraint}
c_\textrm{SDN} > \alpha_\textrm{PCE}+\alpha_\textrm{SDN}
\end{equation}

\par With the above definitions of CapEx (equation \eqref{eq:capex definition}),
OpEx (equation \eqref{eq:opex definition}) and revenue (equation
\eqref{eq:revenue definition}), as applicable for the joint migration to PCE and
SDN, and subject to the associated contraints amongst the various coefficients
(equations \eqref{eq:unmig > mig}-\eqref{eq:constraint}), the payoff function in equation
\eqref{eq:genericPayoffDef}, reduces to,

\begin{multline}
P_i({a \to a^\prime})=
\frac{[(T_{a^\prime}^i)^2-(T_{a}^i)^2] - [c_i(a,a^\prime)+\alpha_i(a^\prime)]\sqrt{T_{a^\prime}^i}+\alpha_i(a)\sqrt{T_a^i}}{c_i(a,a^\prime)\sqrt{T_{a^\prime}^i}}
\end{multline}

\section{Numerical Results} \label{sec:rd}
In this section, we present our simulation framework and the empirical
results to evaluate various aspects of our proposed network migration model.

\subsection{Simulation Model}

For our simulation, we consider a scale-free network of 100 interconnected
network islands, comprising of 39 ``\emph{transit}" islands and 61
``\emph{stub}" islands.  Akin to the terminology used in global Internetworks, a
network island that is not a provider for any other island is called a
\emph{stub island}, while all other islands are called as \emph{transit islands}
\cite{gao}. Stub islands represent the end-users, and hence, the choice of
migration rests only with the transit islands. Our topology was generated using
Barab\'asi and Albert's topology generation algorithm \cite{Albert00}, where the
seed network comprised of 16 fully inter-connected network islands, referred to
as \emph{seed islands} due to their higher resulting connectivity.
In our topology, a node represents a network island and a link represents an
inter-island connection. To comply with policy-aware routing, each edge is
marked as either Customer-to-Provider (C2P) or Peer-to-Peer (P2P). We employ
No-Valley-Prefer-Customer (NVPC) routing to provision connection requests
between two network islands, which comprises of the following two
rules \cite{He12}:
\begin{itemize}
  \item Paths learned from providers or peers are never advertised to other providers or peers.
  \item Paths learned from customers are preferred to the paths learned from peers and providers, and paths learned from peers are preferred to the paths learned from providers, regardless of path length.
\end{itemize}

\par Our simulation concerns with migration
to technologies such as PCE, which are beneficial to a connection request, only
when \emph{all} domains on its path from source to destination, have migrated to
the technology in question. This reflects in our routing algorithm, such as,
while provisioning a connection request, amongst various equi-cost paths, the
source domain prefers a path in which all domains have migrated to PCE.
And if multiple equi-cost, shortest paths exist,
the traffic is uniformly distributed across all such paths, or randomly over
one of these paths, depending on the user preference. We model the incoming connection requests
for each source-destination stub domain pairs as Poisson arrivals. The connection requests once provisioned are assumed to stay the same till the end of simulation. Link capacity is
assumed to be unlimited, since for a given increment in incoming traffic (which
translates to revenue for the host network island), the host network operator
can easily increment the link bandwidth, with minimal effort. This is especially
true since our study is not based on infinitesimal timescales, but of the order
of weeks or months, wherein a domain has the flexibility to increase its link
capacity, subject to incoming requests. All stub-to-stub paths had traffic since
the beginning of the simulation.

\par As a new connection request arrives in a network island, the network
provisions the request and reconsiders its migration choices based on its payoff
function, as defined in Section \ref{sec:model}. This, in turn, leads to its
neighbors reconsidering their respective migration choices, which thus cascades
throughout the network. Finally, on registering a change in the migration
decision of any domain in the network, each domain revises the routes of its
provisioned connections. All the presented results plot average values across 50
traffic profiles (each Poisson distributed), with each traffic profile
replicated 5 times to eliminate any statistical variations.
Paths were precomputed and stored, instead of on-the-fly path computations, as
it significantly improved the simulation run time. The two primary input
preferences to our simulation are (1) \emph{equi-cost routing} ---
when multiple equi-cost paths exist to provision a given connection request, we
consider both possibilities of assigning it to a single random node amongst
them (single-path routing), as well as, that of uniformly distributing the
traffic over all such paths (multipath routing), and, (2) \emph{strategy
estimation approach} --- the approach used by domain to estimate the future
technology deployment in neighboring domains in the process of optimizing its own migration decision; we consider
two approaches for the same, namely, deterministic and probabilistic
approaches, as defined in section \ref{sec:estimation approaches}.

\par We next present our simulation results from various experiments
studying a variety of factors affecting the network \emph{migration profile}.
By `\emph{migration profile}', we mean the progress of the network-wide
migration captured by monitoring the number of migrated nodes throughout the
simulation.
Unless otherwise stated, the parameter values assumed in our simulation are $\eta=1.5$, $c_\textrm{PCE}=0.3$,
$c_\textrm{SDN}=0.4$, $\alpha_\textrm{PCE}=0.1$, $\alpha_\textrm{SDN}=0.2$,
$\alpha_{\overline{\textrm{PCE}}}=0.5$ and
$\alpha_{\overline{\textrm{SDN}}}=0.8$ (though other parameters combinations
were also found to result in similar plots).
The relevant radius for each domain (as defined in section \ref{sec:estimation
approaches}) was set to 5 hops. As can be intuitively expected, the
number of migrants should increase during the simulation, perhaps rapidly in the
beginning, and saturating gradually. This is observed in almost all our case
studies. Further, it is important to note that although the nature of the plots
looks similar \emph{across} the case studies, what is important to note is the
difference in the migration profiles subject to variation in parameters \emph{within}
a case study.

\subsection{Single v/s Double Migration}

\begin{figure}
\begin{center}
\includegraphics[width=0.9\textwidth]{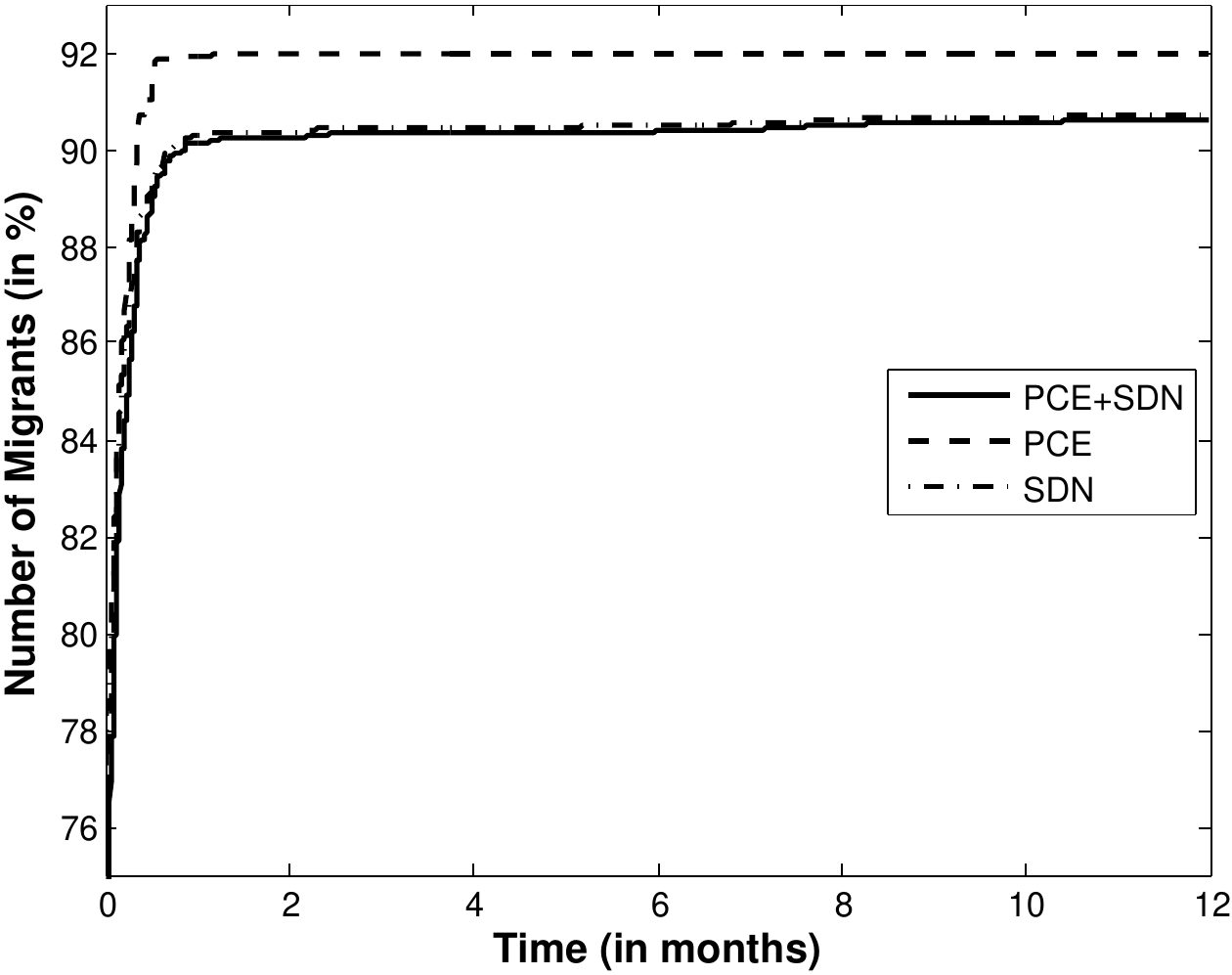}
\caption{Migration profiles of PCE, SDN, and PCE+SDN}
\label{fig:pce-sdn-both}
\end{center}
\end{figure}

\begin{figure}
\begin{center}
\includegraphics[width=0.9\textwidth]{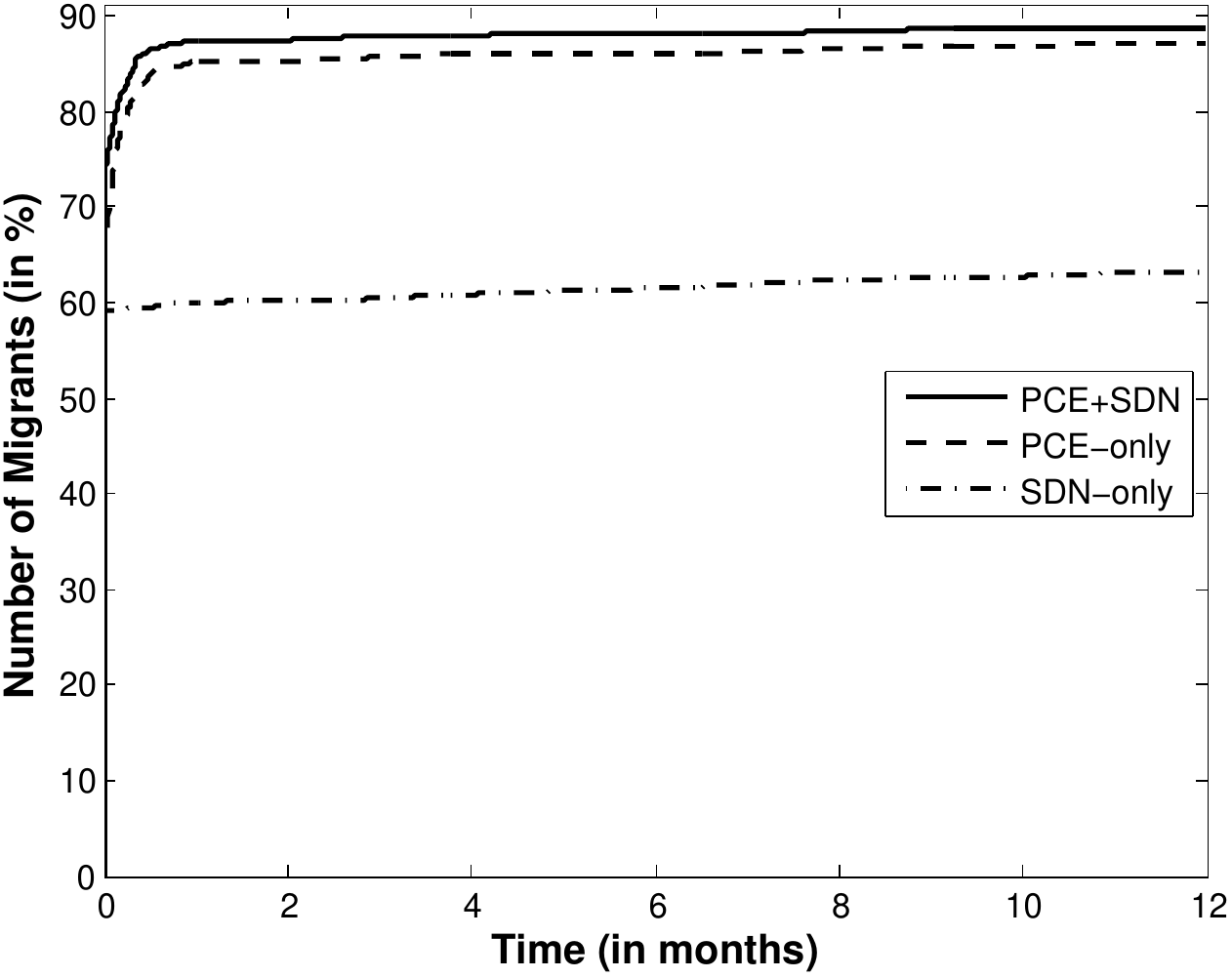}
\caption{Single v/s Double Migrations}
\label{fig:single v/s double migrations}
\end{center}
\end{figure}

\par In this experiment, we study the migration profiles of PCE and SDN, under
varied circumstances.

\par Figure \ref{fig:pce-sdn-both} plots the migration profiles of nodes in the
network to PCE, SDN and PCE+SDN, under probabilistic strategy estimation
approach and multi-path routing preference. Given that we assume migration to
SDN is more expensive than that to PCE (i.e., $\alpha_{PCE}<\alpha_{SDN}$),
Figure \ref{fig:pce-sdn-both} shows that a greater number of nodes migrate to PCE, than SDN, and also that almost every
node that migrates to SDN also migrates to PCE. We observe from Figure
\ref{fig:pce-sdn-both} that none of the domains migrate to SDN, without
migrating to PCE. This demonstrates the fact the benefits derived from SDN are
best exploited in combination with PCE, than by itself.

\par Figure \ref{fig:single v/s double migrations} plots the migration profiles
to PCE and SDN, in three different scenarios, under deterministic strategy
estimation approach and multi-path routing preference.
\emph{PCE-only} plots the PCE migration profile in the network, when only migration to PCE is studied in
isolation, i.e.,  SDN is not considered at all. Similarly, \emph{SDN-only} plots
the SDN migration profile in the network, when only migration to SDN is considered in
isolation, i.e.,  PCE is not studied at all. Finally, \emph{PCE+SDN} plots the
profile of nodes migrating to \emph{both} PCE and SDN, when PCE and SDN
migrations are considered simultaneously. This plot shows that migration to SDN
which is generally small by itself, can be further promoted by joint migration
to PCE, which is more widely accepted, given the complementary relationship
between PCE and SDN. Also, a small increase can be
observed in the PCE migration from \emph{PCE-only} to \emph{PCE+SDN}, thus SDN also has a
small impact in improving the PCE deployment.

\subsection{Early Adopters}

\begin{figure}
\centering
\begin{subfigure}
\centering
\includegraphics[width=0.9\textwidth]{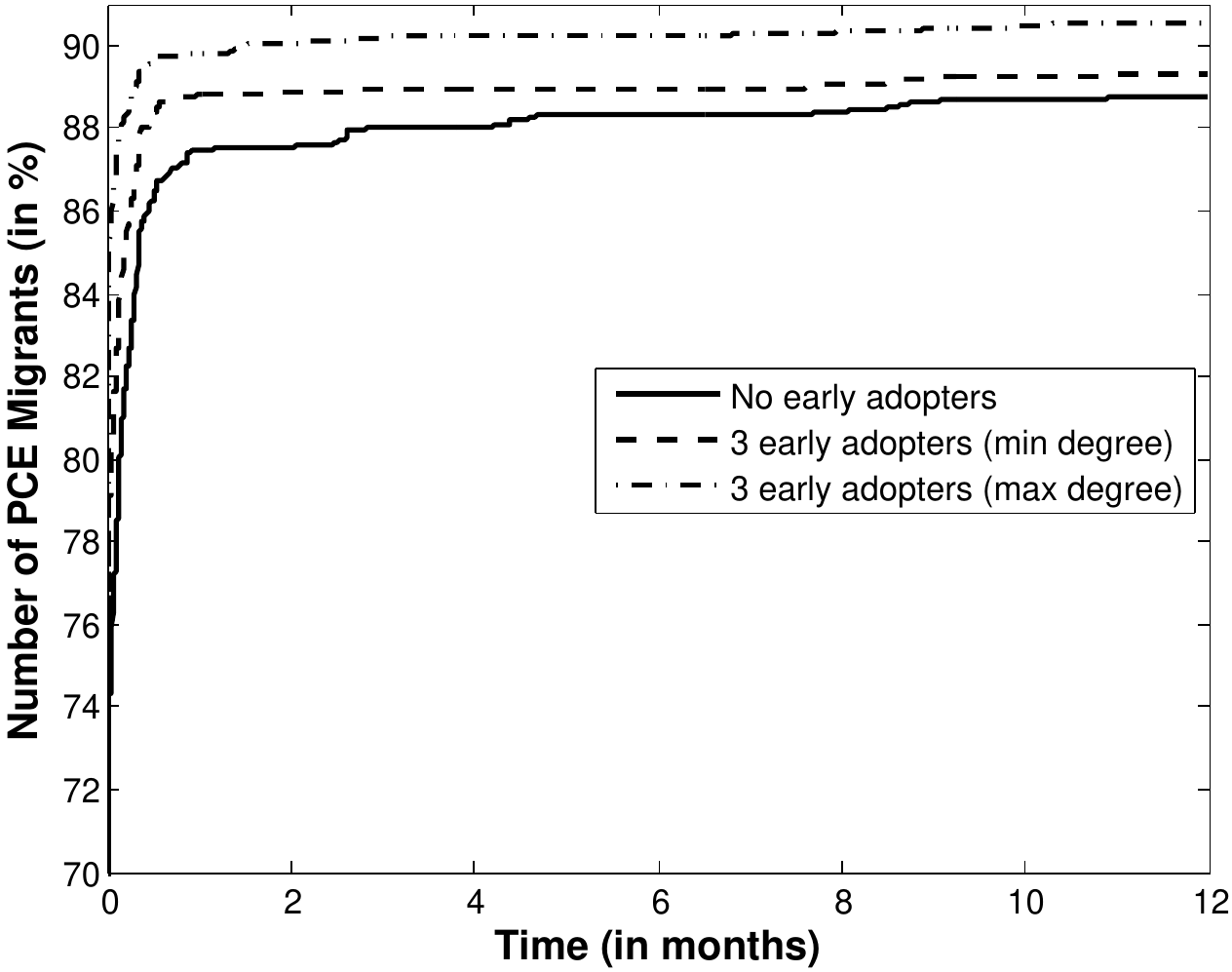}
\end{subfigure}%
~ 
\begin{subfigure}
\centering
\includegraphics[width=0.9\textwidth]{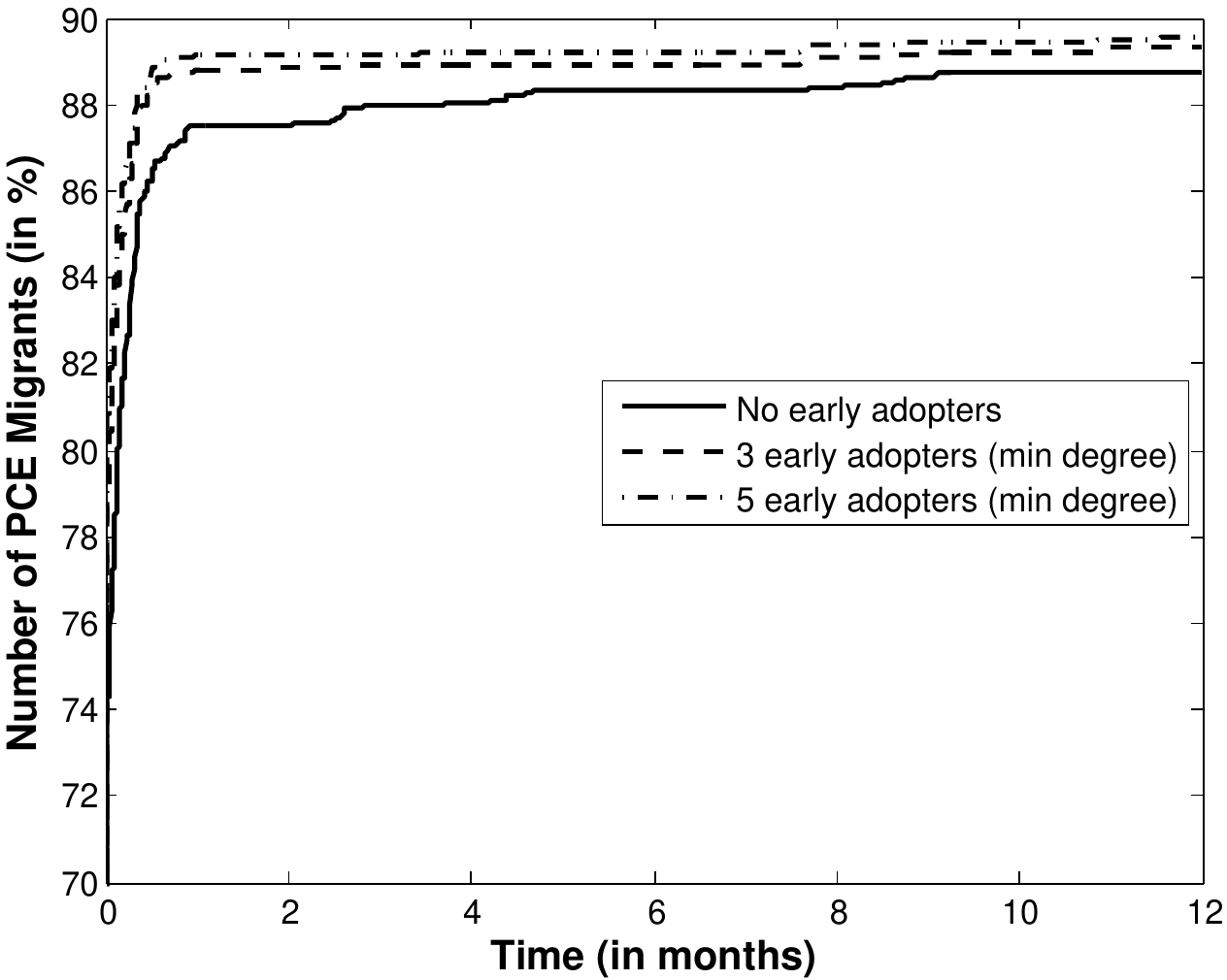}
\end{subfigure}
\caption{Effect of early adopters on PCE migration profile by type (top) and
number (bottom)}
\label{fig:effect of early adopters}
\end{figure}

\par We next study the effect of early adopters on the PCE and SDN migration
profiles in the network, based on the type and number of early adopters. In this
experiment, an early adopter is a network domain that has migrated to PCE since
the beginning of simulation. Early adopters act as the seed for migration in the network,
thereby catalyzing the migration process.

\par Figure \ref{fig:effect of early adopters} (top) plots the effect of type of
PCE early adopters on the PCE migration profile in the network, under
deterministic strategy estimation approach and multi-path routing preference. We
choose the early adopters based on their degree of connectivity in the network.
Figure \ref{fig:effect of early adopters} (top) contrasts the PCE migration
profile in the network given no early adopters, 3 early adopters (amongst the
minimum degree nodes in the network), and 3 early adopters (amongst the maximum
degree nodes in the network). As can be intuitively expected, these plots
suggest that nodes with high degrees, on migrating, have a greater effect in
promoting the network-wide migration profile, than nodes with smaller degrees.
This can be attributed to the fact that a large number of paths pass through
the high-degree nodes in the network. Thus, the migration of a
single high-degree node would affect the migration choices of a large number
of transit nodes, due to its high degree of connectivity.

\par Figure \ref{fig:effect of early adopters} (bottom) plots the effect of
number of PCE early adopters on the PCE migration profile in the network, under
deterministic strategy estimation approach and multi-path routing preference. It
contrasts the PCE migration profile in the network given 0, 3
and 5 early adopters (amongst the minimum degree nodes in the network). As can
be intuitively expected, the plot shows that a higher the number of early
adotpers result in a better migration profile.

\subsection{Cause of Migration}

\begin{figure}
\begin{center}
\includegraphics[width=1\textwidth]{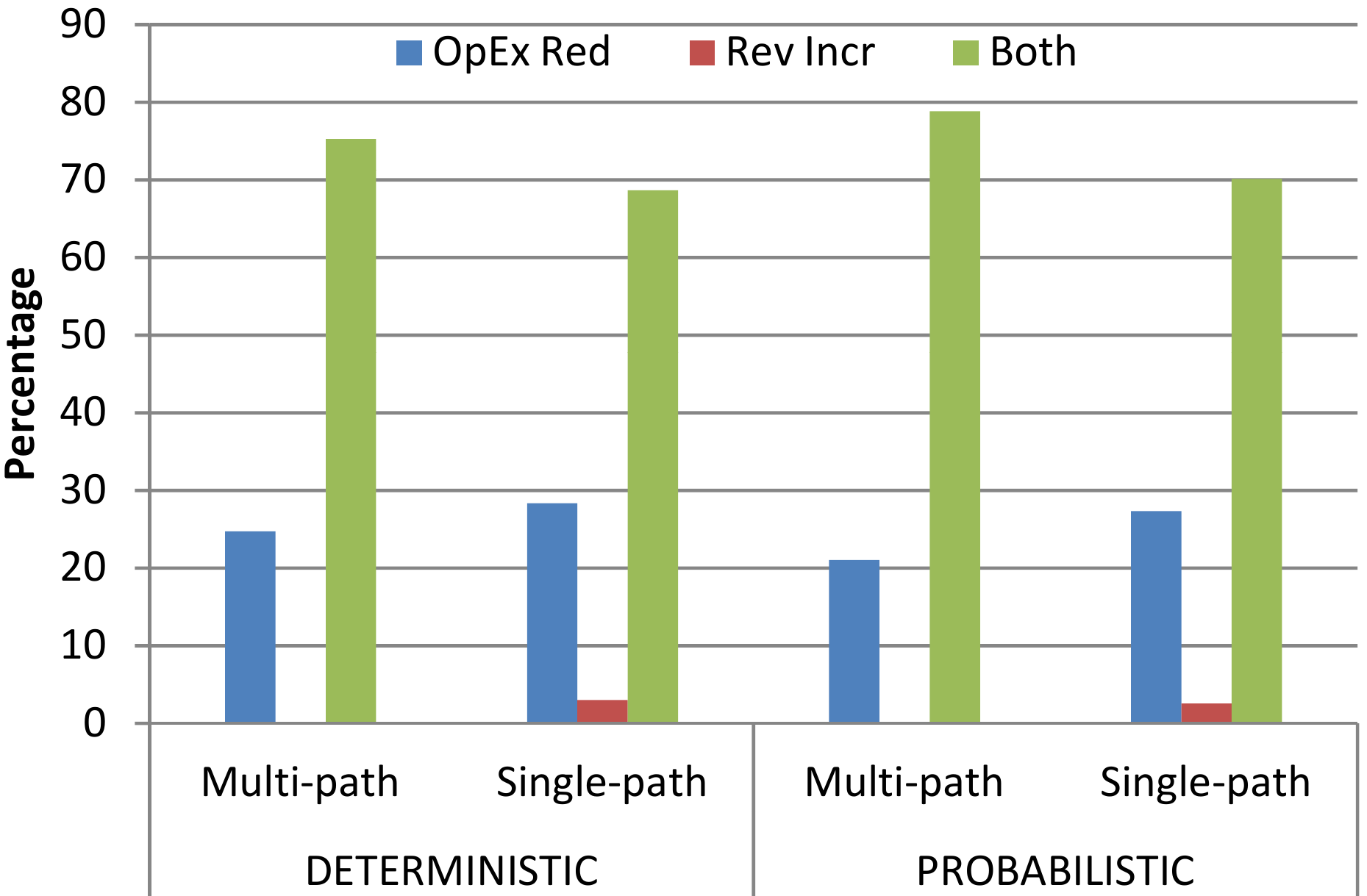}
\caption{Cause of Migration}
\label{fig:cause of migration}
\end{center}
\end{figure}

In this experiment, we study the motivations for transit domains to migrate to
either PCE or SDN or both. As discussed earlier, a transit node migrates either
to reduce its operational expenditures (OpEx), or to increase the traffic
flowing through it (and, in turn its revenue), or both. For every domain that
choose to migrate during the simulation, we monitored them, and categorized
their cause of its migration, amongst (1) exclusive reduction in OpEx, (2)
exclusive increase in traffic (in turn, resulting in an increase in its
revenue), and (3) both (1) and (2). Figure \ref{fig:cause of migration} plots
this data (in percentages) for various combinations of routing choice (single-
or multi-path) and strategy estimation choice (deterministic or probabilistic).
This plot contradicts the common misnomer that a domain migrates
primarily because of a resulting increase in traffic (or revenue). The plot
illustrates an important aspect of migration, which is, a transit node may migrate even if its migration
decision does not result in an increase in traffic (or revenue), but only based
on its OpEx reduction. We observe that a significant fraction of migrations
result exclusively due to decrease in OpEx. Moreover, OpEx reduction proves to
be more important in case of single path routing, than multi-path
routing. Figure \ref{fig:cause of migration} also demonstrates that revenue
increase almost always results in combination with OpEx reduction as a cause of
migration, and rarely in isolation.

\subsection{Effect of Coupling Coefficient}

\begin{figure}
\centering
\begin{subfigure}
\centering
\includegraphics[width=0.9\textwidth]{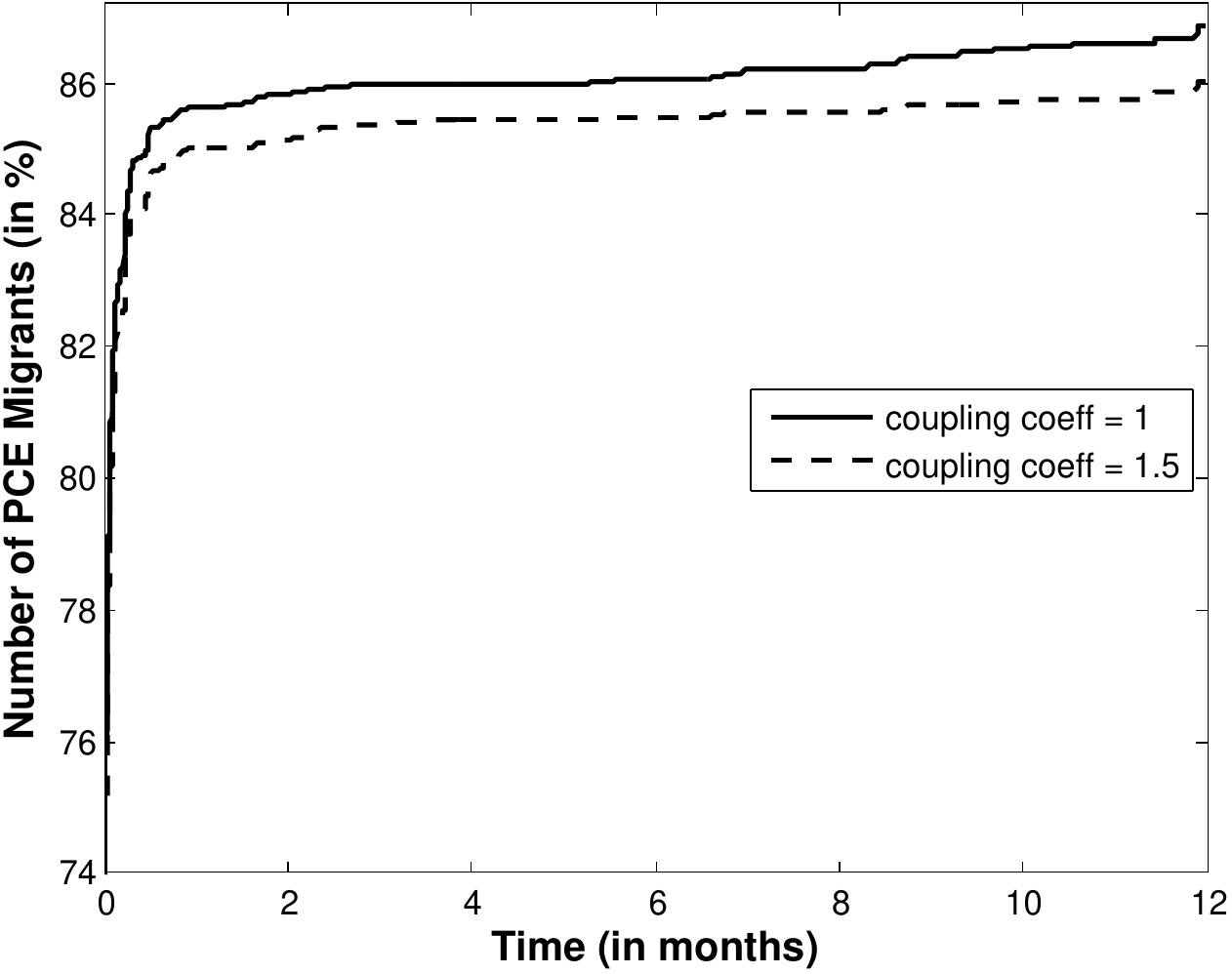}
\end{subfigure}%
~ 
\begin{subfigure}
\centering
\includegraphics[width=0.9\textwidth]{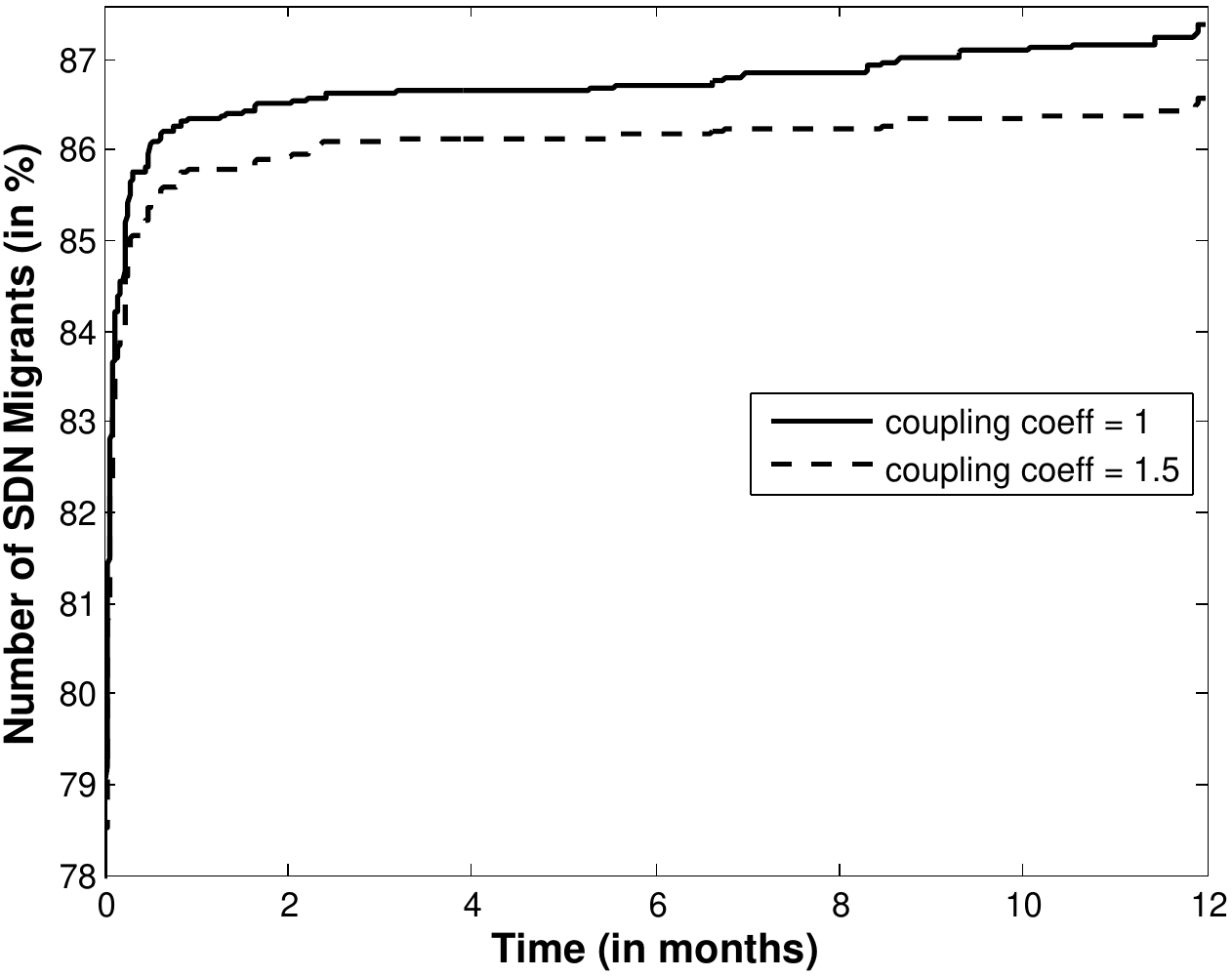}
\end{subfigure}
\caption{Effect of coupling coefficient on PCE (top) and SDN (bottom)
migration profiles}
\label{fig:effect of coupling coeff}
\end{figure}

In this experiment, we study the effect of coupling coefficient on the migration
profile. Figure \ref{fig:effect of coupling coeff} plots the effect of coupling
coefficient on PCE (top) and SDN (bottom) migration profiles in a 150-node
topology with 92 stubs and 58 transits, under deterministic strategy estimation
approach and multi-path routing preference. We observe that the resulting
migration profile is enhanced, when we account for the complementary
relationship between PCE and SDN (coupling coefficient = 1.5) than otherwise
(coupling coefficient = 1).
This is because, when PCE and SDN operate simultaneously in a domain, the
resulting benefits are larger than the sum of benefits derived from PCE and SDN
individually. Thus, domains deploying either PCE or SDN benefit from this
aspect, and also choose to adopt the complementary technology i.e.,  SDN or PCE,
respectively, consequently resulting in a higher number of migrants.

\subsection{Effect of Equi-cost Routing Preferences}

\begin{figure}
\centering
\begin{subfigure}
\centering
\includegraphics[width=0.9\textwidth]{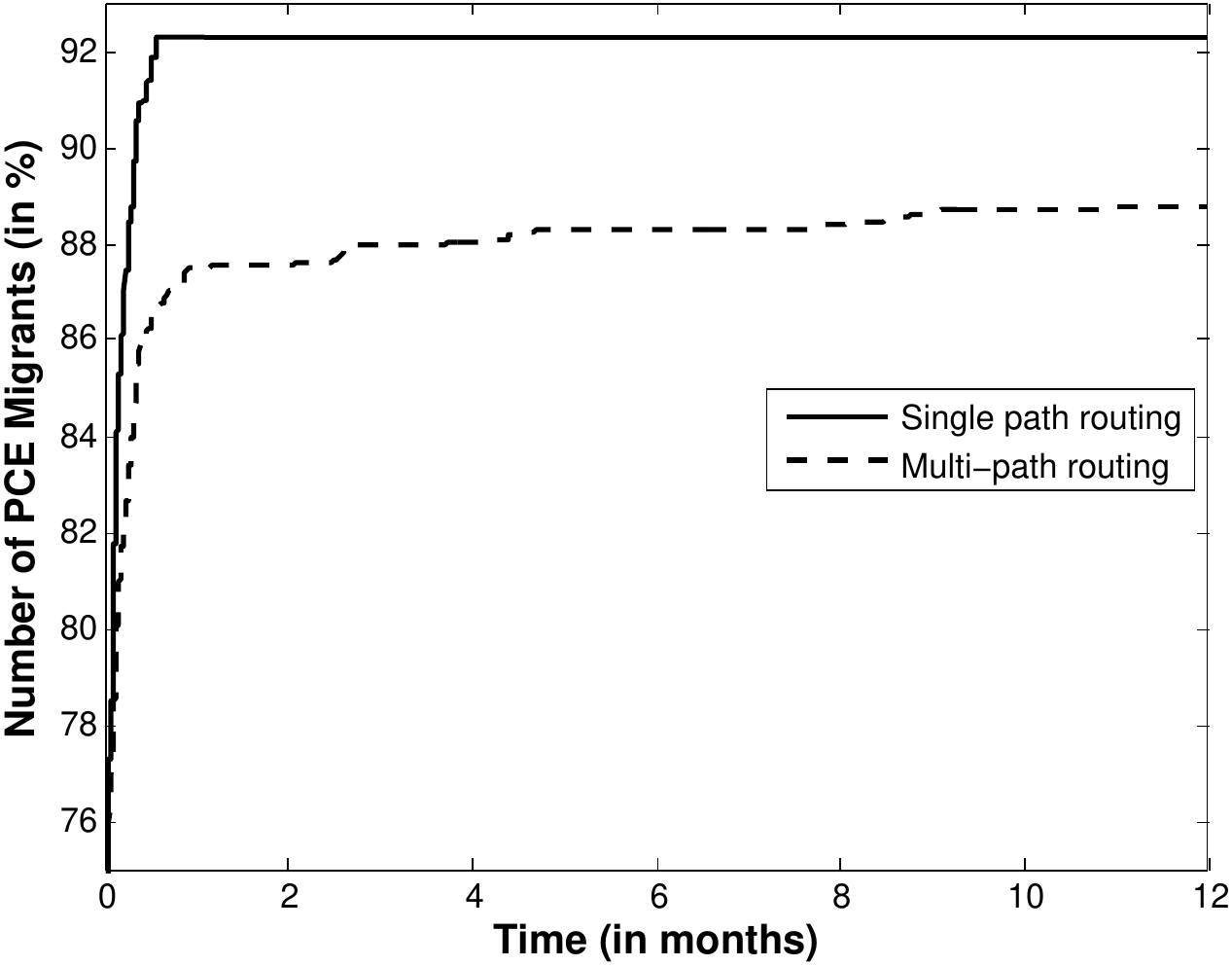}
\end{subfigure}%
~ 
\begin{subfigure}
\centering
\includegraphics[width=0.9\textwidth]{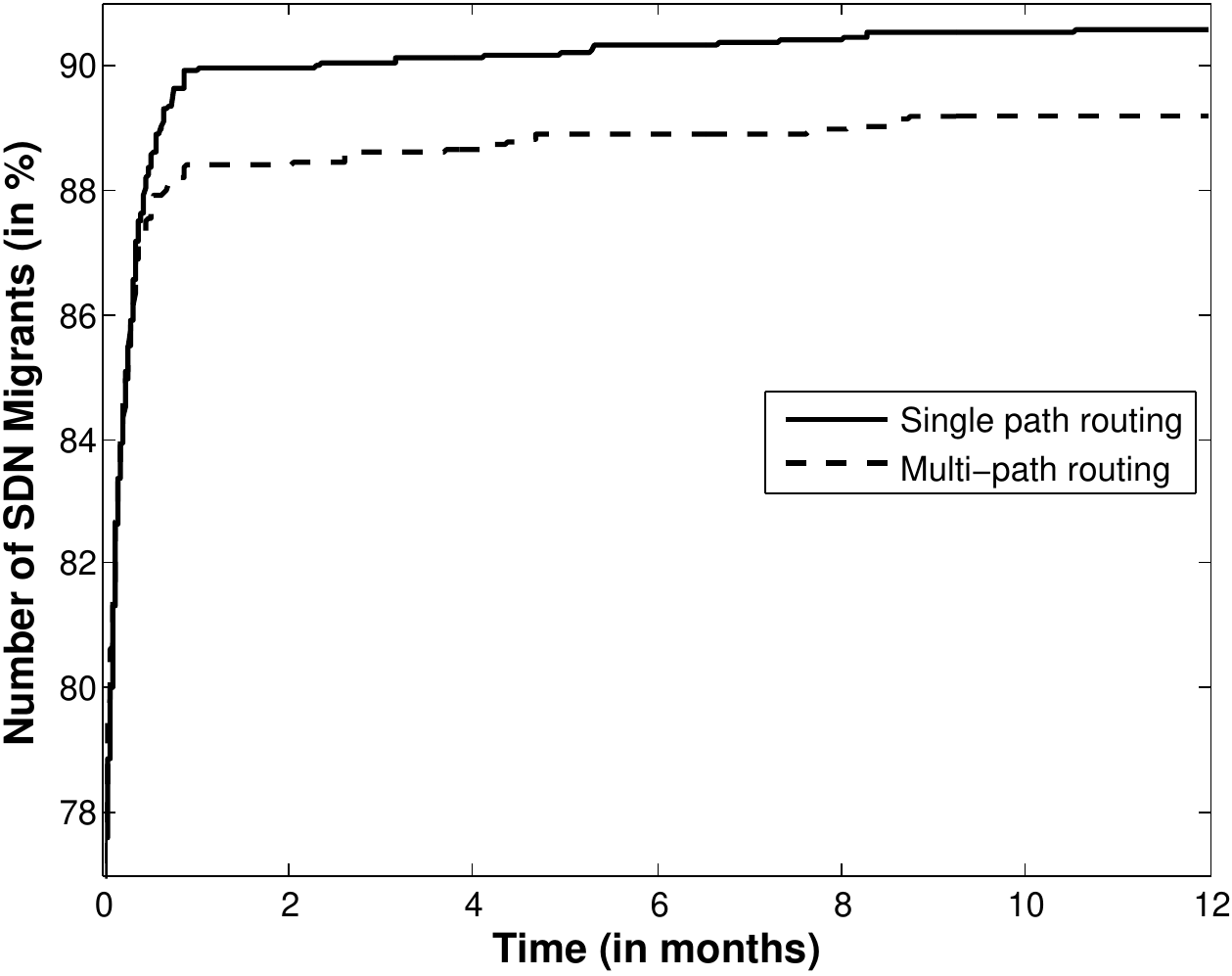}
\end{subfigure}
\caption{Effect of routing choices on PCE (top) and SDN (bottom)
migration profiles}
\label{fig:effect of routing approaches}
\end{figure}

In this experiment, we study the effect of routing choices, when multiple
equi-cost shortest paths exist in the network to provision a user request.
Figure \ref{fig:effect of routing approaches} plots the effect of equi-cost
routing preferences on the PCE (top) and SDN (bottom) migration profile, under
deterministic strategy estimation approach and multi-path routing preference. In
presence of multiple equi-cost shortest paths, we consider the routing choices
of randomly choosing any one of them (single-path routing), or distributing
traffic uniformly across all of them (multi-path routing). As can be observed
from Figure \ref{fig:effect of routing approaches}, the former choice results in an enhanced migration profile than the latter.
This may be attributed to the fact that distributing traffic over multiple paths
reduces the amount of traffic flowing through each such path, thereby lessening
the incentive derived by the intermediate transit nodes from migration.

\subsection{Effect of Network Topology}

\begin{figure}
\centering
\begin{subfigure}
\centering
\includegraphics[width=0.9\textwidth]{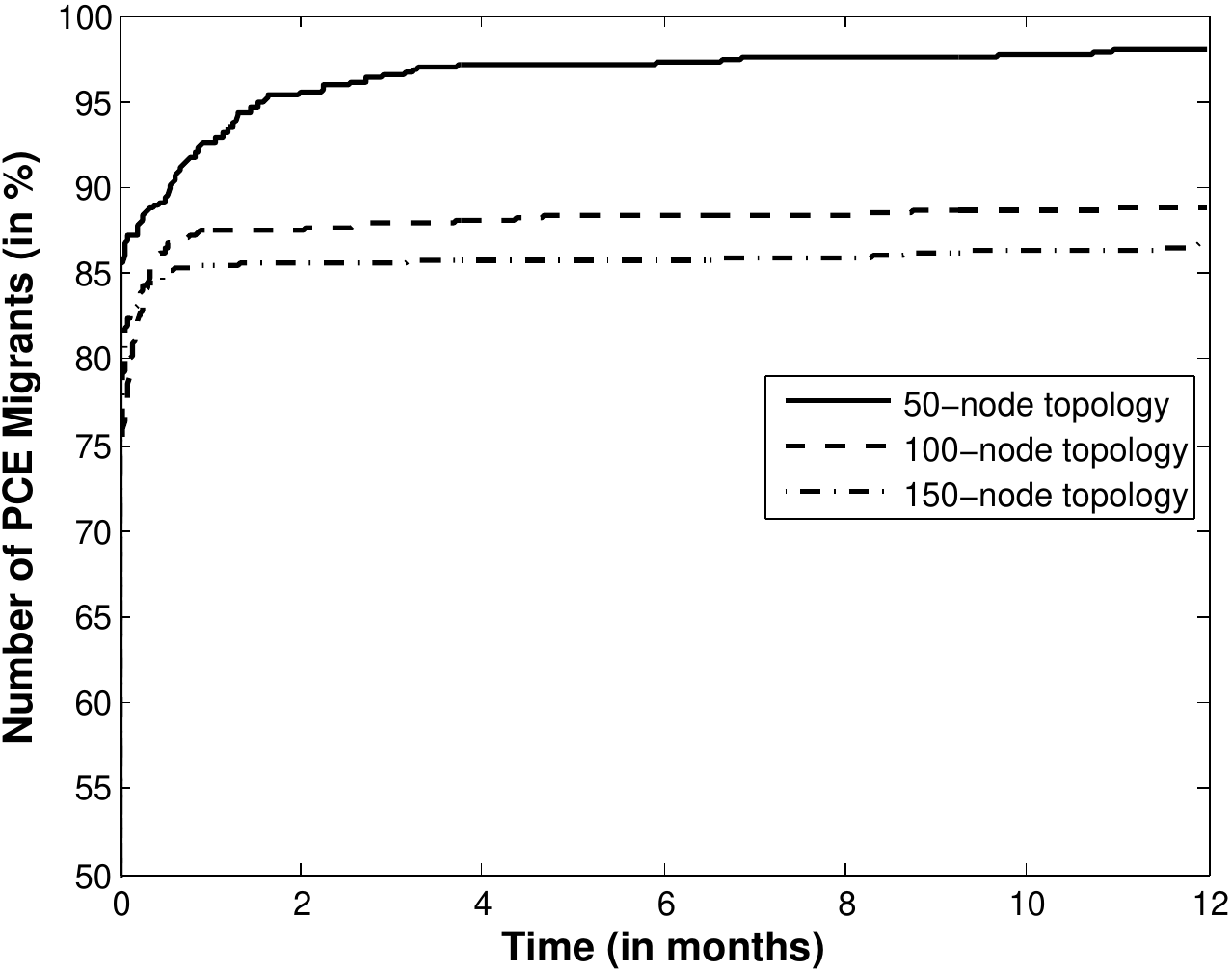}
\end{subfigure}%
~ 
\begin{subfigure}
\centering
\includegraphics[width=0.9\textwidth]{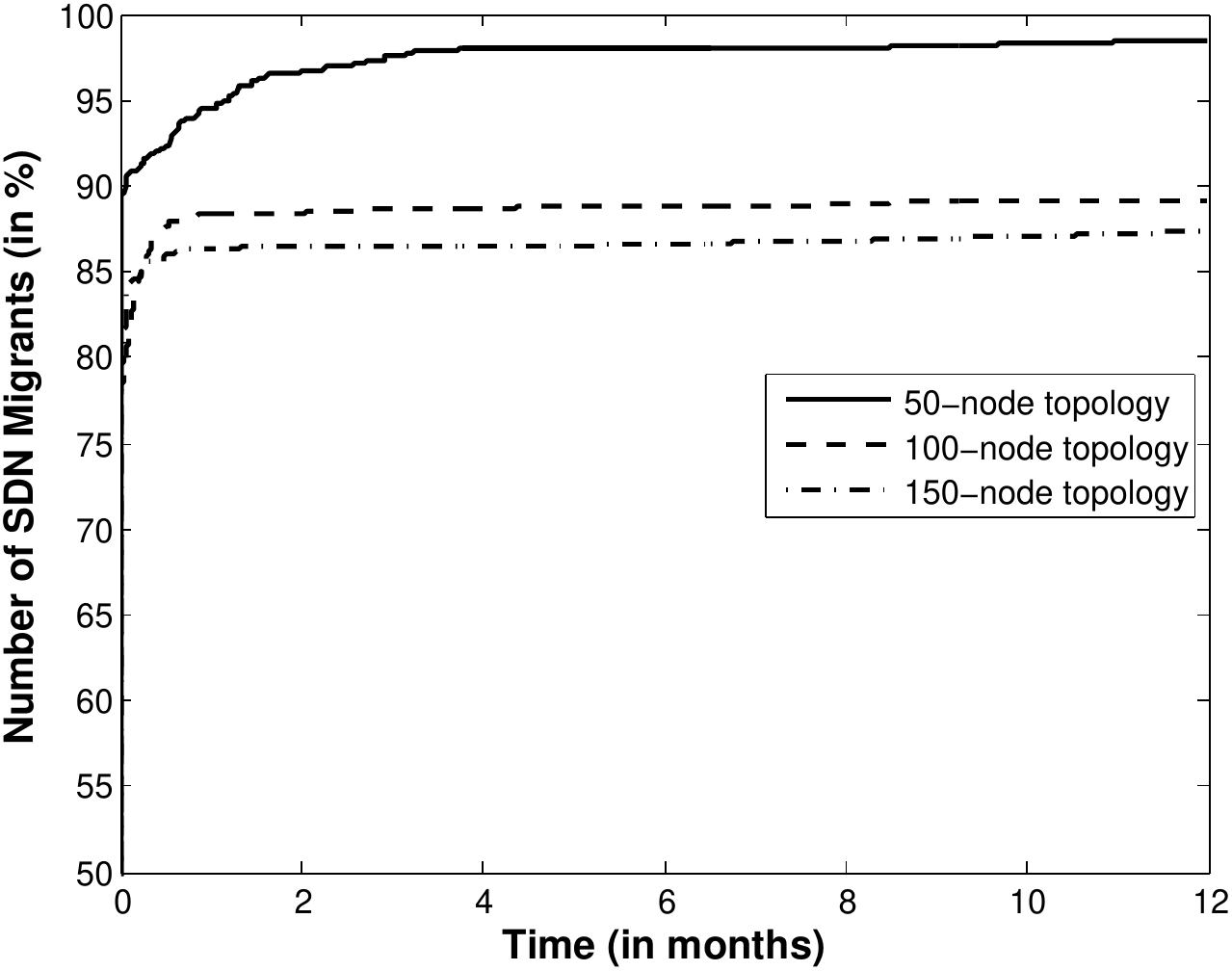}
\end{subfigure}
\caption{Effect of topology size on PCE (top) and SDN (bottom)
migration profiles}
\label{fig:effect of topology size}
\end{figure}

In this section, we discuss the effect of size of the network topology on the
migration profile of a network. In addition to the 100-node topology, we
consider 50- and 150-node topologies, with similar characteristics, in terms of
the fraction of stub/transit nodes in the network, seed network size, degree of
stub nodes, etc. Figure \ref{fig:effect of topology size} plots the percentage
of nodes migrating to PCE (top) and SDN (bottom) migration profiles, under
deterministic strategy estimation approach and multi-path routing preference. We
observe that a larger fraction of nodes migrate in the 50-node topology, than in
the 100-node topology, which in turn has a larger number of migrants than the
150-node topology. This leads us to conclude that for the same set of
parameters, the migration profile is increasingly pronounced in smaller
topologies than larger topologies.

\subsection{Strategy Estimation Approach}

\begin{figure}
\centering
\begin{subfigure}
\centering
\includegraphics[width=0.9\textwidth]{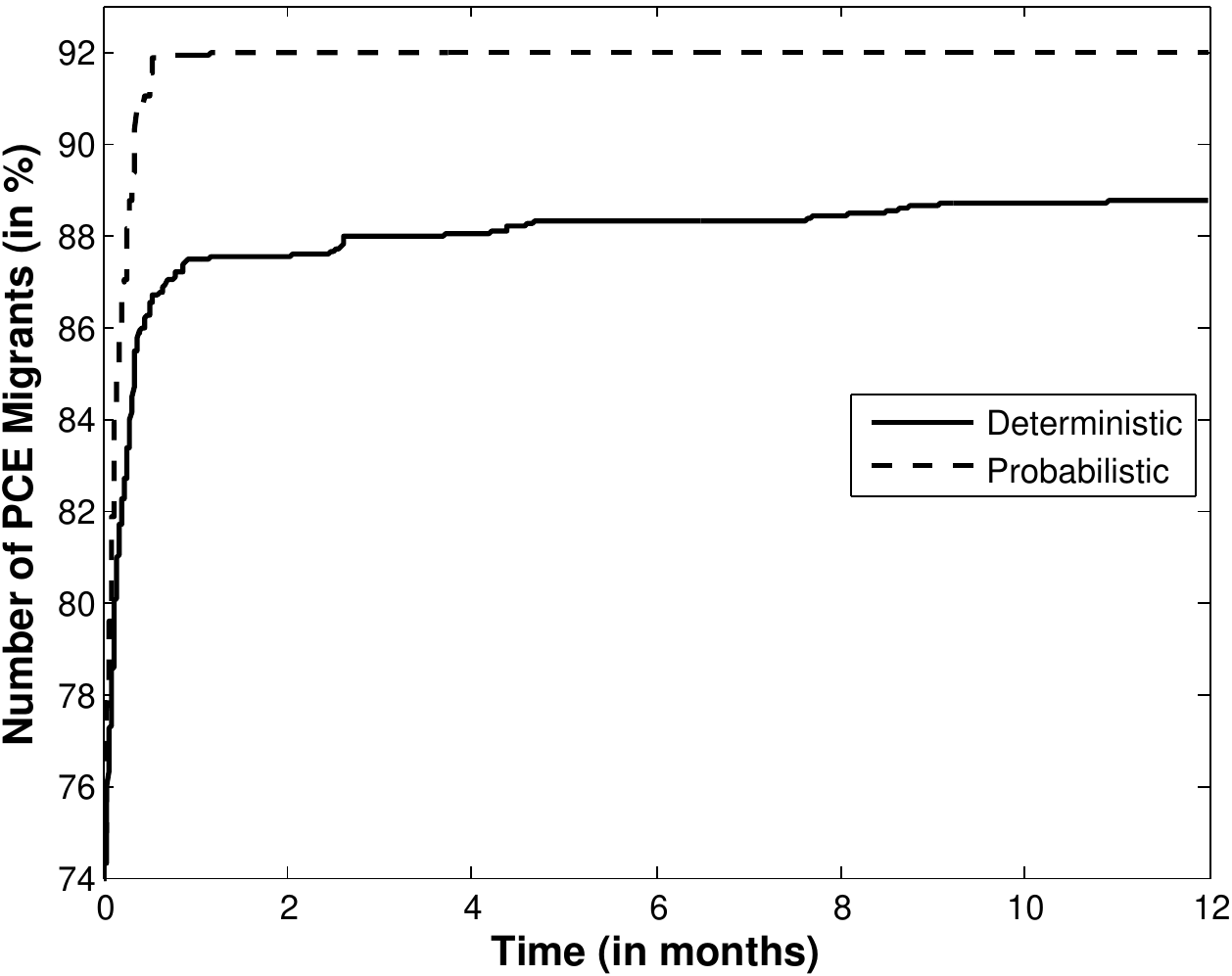}
\end{subfigure}%
~ 
\begin{subfigure}
\centering
\includegraphics[width=0.9\textwidth]{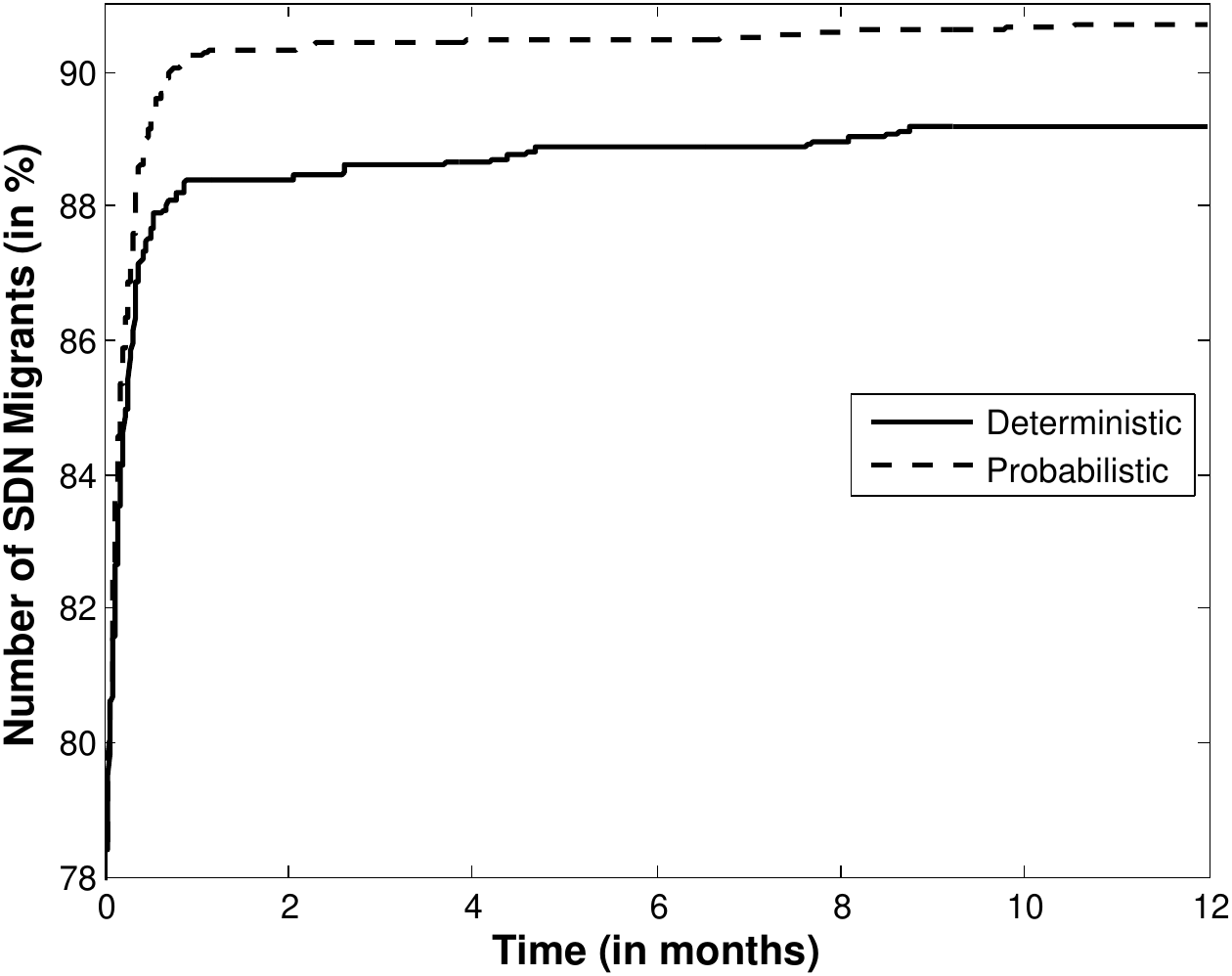}
\end{subfigure}
\caption{Effect of the strategy estimation approaches on PCE (top) and SDN
(bottom) migration profiles}
\label{fig:effect of strategy estimation approach}
\end{figure}

In this experiment, we compare the effect of different strategy estimation
heuristics employed by a domain on the migration profile of the network. Figure
\ref{fig:effect of strategy estimation approach} plots the number of domains
migrating to PCE or SDN over time, when multipath routing is enabled. We observe
that the deterministic approach results in a lesser number of migrants than the
probabilistic approach for both PCE and SDN migrations.

\par This behavior can be explained as follows. As a thumb rule, greater the
number of neighboring migrated domains, greater is the likelihood of a domain to
migrate. In the deterministic and probabilistic approaches, the estimated number
of neighboring migrated domains considered by a domain is \emph{greater than or
equal} to the actual number of migrated domains in the neighborhood. Amongst the
deterministic and probabilisitic approaches, the likelihood of migration of a
node varies with the effective migration coefficient of neighboring nodes as
shown in Figure \ref{fig:deterministic probabilistic curves}. The reader may
note that the area under curves in Figure \ref{fig:deterministic probabilistic
curves} are proportional to the total number of migrations resulting from each
estimation approach. Had the effective migration coefficient of the nodes be uniformly
varying between 0 and 1, both approaches would have resulted in similar
migration profile. However, we observe in our simulation (and can also be
intuitively derived) that the effective migration coefficient varies roughly
between 0 and 0.8, thereby providing the probabilistic approach an upper hand.
As a result, the probabilistic approach results in a greater number of migrants
than that from the deterministic approach.

\par Although more and more transit domains migrate with increasing traffic in
the network, it is important to note that the saturation point of migration is
reached not when \emph{all} transit domains migrate, but at a \emph{lesser} number of
migrants. For example, out of 39 transit nodes in the network, only about 36
migrate at saturation, as seen in Figure \ref{fig:effect of strategy estimation
approach}. This is because of the shortest-path routing between the stub nodes.
Thus, only those transit domains which \emph{lie} on the shortest path(s) between a pair of stub nodes, eventually migrate, whereas,
transit nodes with no stub-to-stub traffic have no incentive in migrating, even
when every other node in its neighborhood may have migrated.

\section{Conclusion} \label{sec:conclusion}

\par In this paper, we proposed an agent-based model to study network migration to
multiple technologies that may be correlated, and applied it to study two
emerging technology frameworks, i.e.,  PCE and SDN. We believe to have advanced the science in the existing agent-based models by considering a few novel  critical factors, including (i) synergistic relationships across multiple
technologies, (ii) reduction in operational expenditures (OpEx) as a reason to
migrate, and, (iii) implications of local network effects on migration
decisions. As is characteristic of
agent-based models, defining the mutual, microscopic interactions between
agents lead to insights about the macroscopic, system-wide behavior, which was
analyzed and demonstrated by our model.

\par The results obtained from our case study suggest that migration to SDN can be
eased by joint migration to PCE, and that the benefits derived from SDN are best
exploited in combination with PCE, than by itself. The case study also showed
that studying migration to related technologies in combination is important than
studying migration to each technology in isolation. The results indicate that
the migration to SDN can be promoted by several factors, namely, (a) in combination with a widely-accepted complementary technology such as PCE, (b) early adopters,
(c) an agent's ability to predict its neighbor's decisions to migrate to either of the technologies.

\par Our future work includes applying our model to study larger
topologies (of the scale of thousands of domains). Also,
multi-vendor, multi-layer network migration scenarios with IP/Optical network
integration is a relevant scenario to investigate. Our model can also be
extended to study inter-relationships between three or more migrating
technologies, which can be explored should a relevant case study emerge. Another important aspect would be to study
the order of migration in a network, i.e.,  ``migration scheduling", showing
which type of nodes should migrate first.


\begin{acknowledgements} \label{sec:ack} 
This work has been supported by the
German Federal Ministry of Education and Research (BMBF) under support code
01BP12300A; EUREKA-Project SASER.
\end{acknowledgements}

\bibliographystyle{IEEEtran}
\bibliography{netmig}

\end{document}